\documentclass[pre,twocolumn,floatfix,hyperlinks]{revtex4-1}

\usepackage{graphicx}
\usepackage{bm}
\usepackage{bbold}
\usepackage{mathtools}
\usepackage{color}
\usepackage{transparent}
\usepackage[colorlinks=true,citecolor=blue]{hyperref}

\DeclareFontFamily{U}{cbgreek}{}
\DeclareFontShape{U}{cbgreek}{m}{n}{
	<-6>    grmn0500
	<6-7>   grmn0600
	<7-8>   grmn0700
	<8-9>   grmn0800
	<9-10>  grmn0900
	<10-12> grmn1000
	<12-17> grmn1200
	<17->   grmn1728
}{}
\DeclareFontShape{U}{cbgreek}{bx}{n}{
	<-6>    grxn0500
	<6-7>   grxn0600
	<7-8>   grxn0700
	<8-9>   grxn0800
	<9-10>  grxn0900
	<10-12> grxn1000
	<12-17> grxn1200
	<17->   grxn1728
}{}

\DeclareRobustCommand{\qoppa}{%
	\text{\usefont{U}{cbgreek}{\normalorbold}{n}\symbol{19}}%
}

\makeatletter
\newcommand{\normalorbold}{%
	\ifnum\pdf@strcmp{\math@version}{bold}=\z@ bx\else m\fi
}
\makeatother

\usepackage{etoolbox}
\makeatletter
\patchcmd{\frontmatter@abstract@produce}
{\vskip200\p@\@plus1fil
	\penalty-200\relax
	\vskip-200\p@\@plus-1fil}
{}
{}
{}
\makeatother

\usepackage{xcolor}
\newcommand{\revision}[1]{{{#1}}}

\begin{document}
	
	\title{Lee-Yang theory of the Curie-Weiss model and its rare fluctuations}
	
	\author{Aydin Deger}
	\affiliation{Department of Applied Physics, Aalto University, 00076 Aalto, Finland}
	\author{Christian Flindt}
	\affiliation{Department of Applied Physics, Aalto University, 00076 Aalto, Finland}
	
	\begin{abstract}
		Phase transitions are typically accompanied by non-analytic behaviors of the free energy, which can be explained by considering \revision{the zeros of the partition function in the complex plane of the control parameter and their approach to the critical value on the real-axis as the system size is increased}. Recent experiments have shown that partition function zeros are not just a theoretical concept. They can also be determined experimentally by measuring fluctuations of thermodynamic observables in systems of finite size. Motivated by this progress, we investigate here the partition function zeros \revision{for} the Curie-Weiss model of spontaneous magnetization using our recently established cumulant method. Specifically, we extract the leading Fisher and Lee-Yang zeros of the Curie-Weiss model from the fluctuations of the energy and the magnetization in systems of finite size. \revision{We develop a finite-size scaling analysis of the partition function zeros, which is valid for mean-field models, and which allows us to extract both the critical values of the control parameters and the critical exponents, even for small systems that are away from criticality}. \revision{We also show} that the Lee-Yang zeros carry important information about the rare magnetic fluctuations \revision{as} they allow us to predict many essential features of the large-deviation statistics of the magnetization. \revision{This finding may constitute a profound connection between Lee-Yang theory and large-deviation statistics.}
		
	\end{abstract}
	
	\maketitle
	
	\section{Introduction}
	
	In their seminal works on statistical physics, Lee and Yang developed a rigorous theory of phase transitions by considering the zeros of the partition function in the complex plane of the control parameter \cite{Lee1952,Yang1952a,Blythe2002,Bena2005}. Specifically, they showed how the partition function zeros with increasing system size will move onto the real value of the control parameter for which a phase transition occurs. These ideas provide a \revision{detailed} understanding of phase transitions in a wide range of many-body systems from such diverse fields as protein folding \cite{Lee2013,Lee2013b,Deger2018}, percolation \cite{Arndt2001,Dammer2002,Krasnytska2015,Krasnytska2016}, and Bose-Einstein condensation \cite{Borrmann2000,Mulken2001,Dijk2015,Gnatenko_2017,Gnatenko2017}. It has also been realized that Lee-Yang theory applies not only to equilibrium phase transitions; the framework is also useful to describe non-equilibrium situations such as space-time phase transitions in glass formers~\cite{Biroli2013,Flindt2013,Hickey2013,Hickey2014} or dynamical phase transitions in quantum many-body systems after a quench~\cite{Heyl2013,Zvyagin2016,Heyl2018}. Moreover, Lee-Yang theory has been extended to quantum phase transitions~\cite{Lamacraft2008}.
	
	On top of this, several recent works have shown that partition function zeros are not just a theoretical concept~\cite{Flindt2013,Wei2012,Wei2014,Kuzmak_2019,Kuzmak_2019b,Krishnan2019}. They can also be experimentally determined in engineered nano-structures~\cite{Binek1998,Peng2015,Brandner2017,Flaschner2018}. In one approach, the partition function zeros are found by measuring the fluctuations of thermodynamic observables in systems of finite size \cite{Flindt2013,Brandner2017}. This approach has been used to extract the dynamical Lee-Yang zeros of an open quantum system by measuring the full counting statistics of tunneling events~\cite{Maisi2014,Brandner2017}. Theoretically, the method has been applied to space-time phase transitions in glass formers and open quantum systems~\cite{Flindt2013,Hickey2013,Hickey2014} as well as equilibrium phase transitions in molecular zippers~\cite{Deger2018} and the Ising model \cite{Deger2019}. The partition function zeros are extracted from the high cumulants of a fluctuating observable, such as energy or magnetization, which can be measured (or simulated) without knowing the partition function~\cite{Flindt2013,Hickey2013,Hickey2014,Deger2018,Deger2019}. The cumulant method appears to be very general; however, further work is needed to fully understand its scope  and potential applications.
	
	\begin{figure}
		\centering
		\includegraphics[width=1\columnwidth]{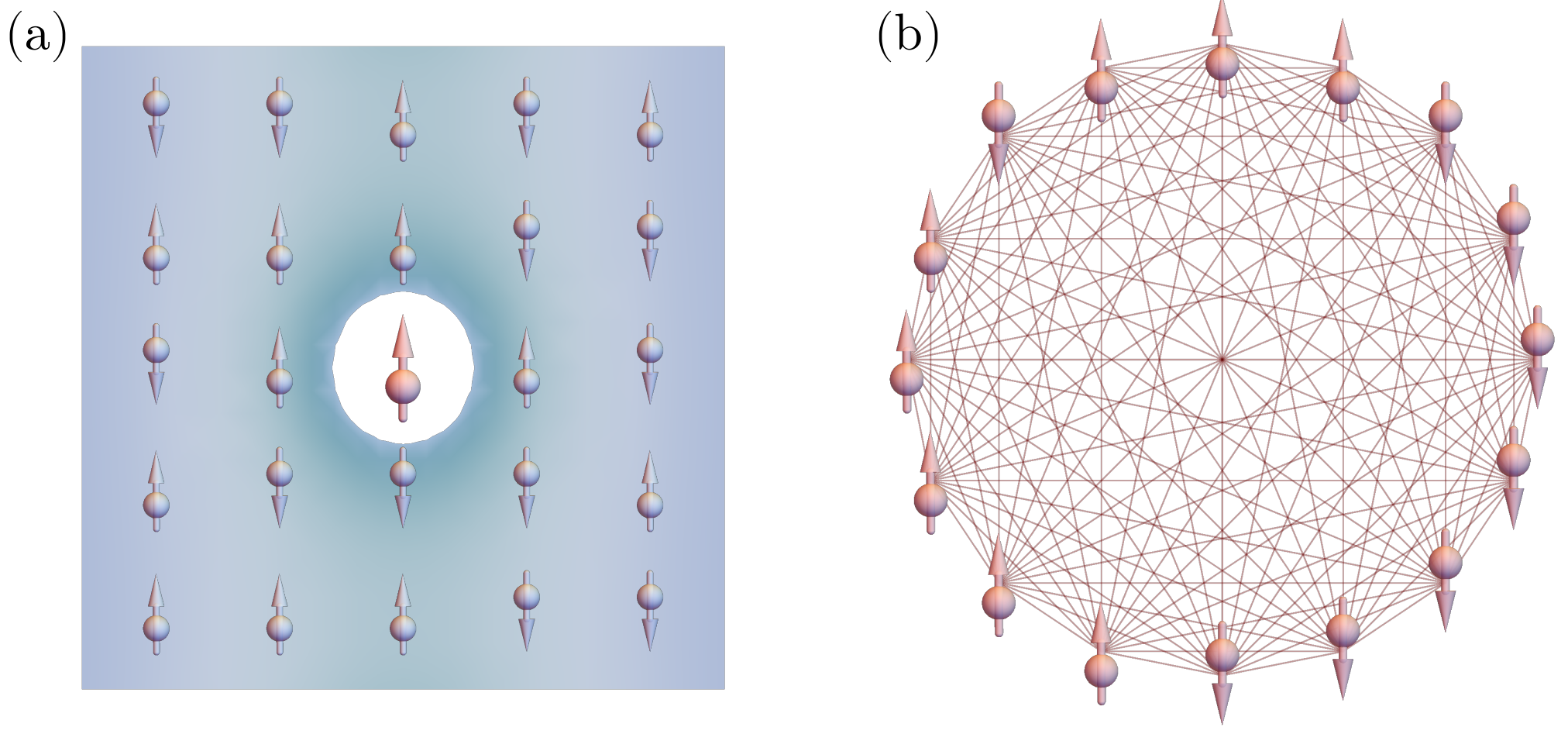}
		\caption{Curie-Weiss model and its graph representation. (a) The Curie-Weiss model follows from a mean-field approximation of the Ising model in which each spin is coupled to the average magnetic field produced by all spins in the lattice. (b)~As a result of the mean-field approximation, the Curie-Weiss model can be represented by a complete graph showing how all spins interact with each other as indicated by lines.}
		\label{fig:fig1}
	\end{figure}
	
	The purpose of this work is to investigate the Curie-Weiss model of spontaneous magnetization using the cumulant method. The model follows from a mean-field approximation of the Ising model (Fig.~\ref{fig:fig1}) and it is attractive because it can be treated with a combination of analytical and numerically exact methods without resorting to simulations \cite{Gaspard_2012,friedli_velenik_2017}. As such, it provides an important testbed for the cumulant method, and it allows us to benchmark our results against numerically exact calculations and thereby improve our understanding of the method. In doing so, we also uncover several interesting properties of the Curie-Weiss model, including its scaling behavior close to criticality and the large-deviation statistics of the rare fluctuations of the magnetization. In particular, we extract the leading partition function zeros from the fluctuations of the energy and the magnetization in the system at finite size. We can then determine their convergence points in the thermodynamic limit based on a finite-size scaling analysis, \revision{which we develop for the Curie-Weiss model}. In this way, we extract both the critical exponents and the critical point of the system, even if the control parameters are fixed, and the system is away from criticality. Finally, we show how the partition function zeros encode important information about the rare fluctuations of the magnetization, and they allow us to make predictions of many essential features of the large-deviation statistics. \revision{This finding suggests that a profound connection between Lee-Yang theory and large-deviation statistics may exist.}
	
	The rest of our work is organized as follows. In Sec.~\ref{sec:CWm}, we  recall how the Curie-Weiss model follows from a mean-field approximation of the Ising model, and we describe some known results about its thermodynamic properties. In Sec.~\ref{sec:3}, we discuss the Lee-Yang theory of phase transitions, including the partition function zeros in the complex plane of the control parameter and their approach to the critical value in the thermodynamic limit. We also introduce our cumulant method, which we use to extract the leading partition function zeros from the fluctuations of thermodynamic observables in the Curie-Weiss model. In Sec.~\ref{sec:scaling}, \revision{we develop a scaling analysis of the partition function zeros, which is important in the following sections, where we determine the critical exponents and the convergence points of the partition function zeros with increasing system size. Here, we need to pay special attention to the mean-field nature of the Curie-Weiss model, which changes the scaling analysis compared to the Ising model in two and three dimensions.} 
	
	Sections \ref{sec:fisher} and \ref{sec:lee-yang} contain our main results. First, we show how the \revision{Fisher zeros in the complex plane of the inverse temperature} together with the critical exponent of the heat capacity can be extracted from the energy fluctuations, which in principle can be either measured or simulated. Importantly, the fluctuations can be obtained at a single fixed temperature, which may be below or above the critical temperature. As such, our method may enable experimental investigations of phase transitions that would otherwise be hard to reach, for instance at very low temperatures. Second, we extract the Lee-Yang zeros in the complex plane of the magnetic field from the fluctuations of the magnetization. Again, based on finite-size scaling arguments, we can extract the critical exponent of the magnetization as well as the convergence points of the Lee-Yang zeros in the thermodynamic limit. In Sec.~\ref{sec:LDF}, \revision{we connect Lee-Yang theory to the field of large-deviation statistics by showing}  that the magnetic field zeros carry important information about the rare fluctuations of the magnetization.  Section~\ref{sec:conclusions} contains a summary, our conclusions and an outlook for future work.
	
	\section{Curie-Weiss model}
	\label{sec:CWm}
	
	We consider \revision{the Curie-Weiss model, which follows from a mean-field approximation of the Ising model, describing} a $d$-dimensional lattice of interacting classical spins in an external magnetic field $h$~\cite{Gaspard_2012,friedli_velenik_2017}. The total energy corresponding to a given spin configuration, $\{\sigma_i\}$, \revision{reads}
	\begin{equation}
		U_{\rm Ising}\left(\{\sigma_i\}\right)  = -J\sum_{\langle i,j\rangle}\sigma_i\sigma_j-h\sum_{i}\sigma_i,
	\end{equation}
	where the spin on site $i$ is defined as $\sigma_i=\pm1$, the angular brackets indicate summation over neighboring spins, and~$J$ is the strength of the spin interactions. In one and two dimensions, the Ising model can be solved analytically, while the problem in three dimensions remains unsolved. In higher dimensions, the solution is eventually given by the mean-field result that we discuss here. 
	
	To arrive at the mean-field description of the Ising model, we replace the sum over neighboring spins, $\sum_{\langle i,j\rangle}\sigma_i\sigma_j$, by the approximation, $\frac{z}{2N} \sum_{i,j}\sigma_i\sigma_j$, where $z$ is the number of neighbors for each spin (the coordination number), and $N$ is the total number of lattice sites. The factor of $1/2$ is included to avoid double-counting, and the factor of $1/N$ ensures that the total energy is extensive in the thermodynamic limit of large lattices. In the following, we absorb the coordination number into the interaction strength by redefining it as $Jz\rightarrow J$, which stays finite in the limit $z\rightarrow\infty$, where the mean-field solution becomes exact. With these approximations, the energy of the Curie-Weiss model becomes~\cite{Gaspard_2012,friedli_velenik_2017}
	\begin{equation}
		U\left(\{\sigma_i\}\right) =-\frac{J}{2 N} \sum_{i,j=1}^{N} \sigma_{i} \sigma_{j}-h \sum_{i=1}^{N} \sigma_{i},
	\end{equation}
	which can also be written as
	\begin{equation}
		U\left(\{\sigma_i\}\right) =-\frac{J}{2 N} M^2-h M 
	\end{equation}
	in terms of the total magnetization $M=\sum_{i=1}^{N} \sigma_{i}$. 
	
	The Curie-Weiss model can be described by a complete graph as illustrated in Fig.~\ref{fig:fig1}b. The model is attractive as it is analytically tractable, but it comes with the caveat that even spins that are far apart interact with each other. To review the phase behavior of the Curie-Weiss model, we consider the partition function
	\begin{equation}
		Z(\beta,h)=\sum_{\{\sigma_i\}} e^{ -\beta U\left( \{\sigma_i\} \right) }.
	\end{equation}
	where $\beta=1/(k_B T)$ is the inverse temperature.
	The partition sum can be written as
	\begin{equation}
		\label{eq:partition_function}
		Z(\beta, h )=\sum_{n=0}^{N} \binom{N}{n} e^{\frac{\beta J}{2 N}(N-2 n)^{2}+ \beta h\left(N - 2 n\right)},
	\end{equation} 
	where the magnetization,
	$M=N-2n$, is given by the number $n$ of spins pointing down, and the binomial coefficient yields the number of such spin configurations.
	
	For large system sizes, we can replace the sum by an integral and write the partition function as~\cite{Gaspard_2012,friedli_velenik_2017}
	\begin{equation}
		Z\left(\beta, h \right)\simeq\frac{N}{2} \int_{-1}^{1} dm ~ e^{-\beta N g(m)},
	\end{equation}
	where $m=M/N$ is the magnetization per site, and we have obtained the effective free energy per site
	\begin{equation} \label{eq:effFree}
		\begin{split}
			g(m)&\simeq -h m - \frac{J m^2}{2} \\
			& +\beta^{-1}\left[\frac{1+m}{2} \ln\left( \frac{1+m}{2} \right)  +\frac{1-m}{2} \ln\left( \frac{1-m}{2}\right)\right] 
		\end{split}
	\end{equation}
	using Stirling's approximation of the binomial coefficient. The integral representation of the partition function is useful as it enables a saddle-point approximation for large lattice sizes, where we can express it as 
	\begin{equation}
		Z\left(\beta, h \right) \approx \sum_i e^{-\beta N g(m_{i})}
	\end{equation}
	in terms of the stable saddle points, $m_{i}=m_{i}(\beta,h)$, which minimize the effective free energy, $g'(m_{i})=0$, and yield the equilibrium magnetization by solving the equation
	\begin{equation}
		\label{eq:averageMag}
		m_{i}=\tanh \left( \beta h + \beta Jm_{i} \right).
	\end{equation}
	The number of solutions depends on the temperature and the magnetic field. Without a magnetic field, there is only one solution, $m_0=0$, at high temperatures, $\beta < 1/J$. In this case, the free energy $F=U-TS$ is dominated by the entropy $S$, and the system is in the disordered phase with no spontaneous magnetization. However, as the temperature is lowered, two nontrivial and stable solutions at $m_{\pm 1}\simeq\pm \sqrt{3(1-1/\beta J)}/\beta J $ develop smoothly from the  one at $m_{0}=0$, which becomes unstable. Thus, at low temperatures, where the free energy is dominated by the internal energy,  the system is in an ordered phase with non-zero spontaneous magnetization. Hence, the Curie-Weiss model undergoes a continuous phase transition at the Curie temperature
	\begin{equation}
		\beta_c=1/J.
	\end{equation}
	Below the Curie temperature, the system also exhibits a first-order phase transition as a function of the magnetic field, since the average magnetization exhibits a discontinuity at zero magnetic field, $h_c=0$. Finally, in the thermodynamic limit, the free energy per site becomes
	\begin{align}
		f(\beta,h)=\frac{F(\beta,h)}{N} =-\frac{\ln Z}{\beta N}&\approx \min_m \{g(m)\}.
	\end{align}
	We note that the free energy can also be found using a Hubbard-Stratonovich transformation \cite{Salinas2001,friedli_velenik_2017}.
	
	\section{THE CUMULANT METHOD \label{sec:3}}
	
	Phase transitions, such as the one described above, are signaled by singularities in the free energy. To understand how such non-analytic behaviors can develop from the partition function, which is analytic for finite systems, Lee and Yang investigated the zeros of the partition function in the complex plane of the external control parameter \cite{Lee1952,Yang1952a,Blythe2002,Bena2005}. For the Curie-Weiss model, the control parameter could be the inverse temperature or the magnetic field ($\beta$ or $h$), and we denote it by $q$. Since the partition sum is an entire function, it can be factorized as 
	\begin{equation}
		\label{eq:parZeros}
		Z(q)=Z(0)e^{cq} \prod_k\left(1-q/q_k\right),
	\end{equation}
	where $q_k$ are the zeros in the complex plane of the control parameter, and $c$ is a constant. The zeros cannot be real, since the partition function is a sum of exponentials. Moreover, they come in complex conjugate pairs, $q_k$ and $q^*_k$, since the partition sum is real for real values of the control parameter. If the control parameter is the magnetic field, the partition function zeros are called Lee-Yang zeros, while for the inverse temperature, they are typically referred to as Fisher zeros. In the works by Lee and Yang, they showed that the partition function zeros in the thermodynamic limit approach the real value of the control parameter for which a phase transition occurs. These ideas now form a rigorous foundation of phase transitions in statistical physics. However, for a long time, partition function zeros were considered a purely theoretical concept, and only recently they have been experimentally determined~\cite{Binek1998,Peng2015,Brandner2017,Flaschner2018}.
	
	In one approach, the partition function zeros are detemined from the fluctuations of thermodynamic observables in small systems~\cite{Flindt2013,Hickey2013,Hickey2014,Deger2018,Deger2019}. Here we briefly discuss this method before applying it to the Curie-Weiss model. We first note that the free energy (or the logarithm of the partition sum) delivers the cumulants of the observable $\Phi$ that is conjugate to the control parameter,
	\begin{equation}
		\label{eq:qCumulants0}
		\langle \! \langle \Phi^n \rangle \! \rangle=\partial_{q}^n \ln Z.
	\end{equation}
	If the control parameter is the magnetic field, the conjugate variable is the magnetization, and if it is the inverse temperature,  the conjugate variable is the energy. 
	
	Now, using the factorization of the partition function in terms of its zeros, we readly find the relation~\cite{Flindt2013,Deger2018,Deger2019}
	\begin{equation} 
		\label{eq:qCumulants}
		\langle\!\langle \Phi^{n}\rangle\!\rangle=- \sum_{k} \frac{(n-1)!}{(q_k-q)^n}, \quad n>1.
	\end{equation} 
	between the cumulants and the partition function zeros. Importantly, the high cumulants are mainly determined by the complex conjugate pairs of zeros $q_{\rm o}$ and $q_{\rm o}^*$ that are closest to the real axis, since they dominate the sum. The contributions from sub-leading zeros are suppressed with the distance to $q$ and the cumulant order~$n$. The high cumulants can then be approximated by only including the leading zeros in the sum and writing them as
	\begin{equation} 
		\label{eq:qCumulantsapp}
		\langle\!\langle \Phi^{n}\rangle\!\rangle\simeq-(n-1)! \frac{2\cos[n\arg(q_{\rm o}-q)]}{|q_{\rm o}-q|^n}, \quad n\gg1.
	\end{equation} 
	Finally, we can determine the leading partition function zeros by inverting this expression as~\cite{Flindt2013,Deger2018,Deger2019}
	\begin{equation}
		\label{eq:QMethod}
		\begin{bmatrix}
			2~\mathrm{Re}[q_{\rm o}-q]\\
			|q_{\rm o}-q|^2
		\end{bmatrix}=\begin{bmatrix}
			1& -\frac{\mathsf{\mu}_{n}^{(+)}}{n}\\
			1& -\frac{\mathsf{\mu}_{n+1}^{(+)}}{n+1}
		\end{bmatrix}^{-1}
		\begin{bmatrix}
			(n-1) \mathsf{\mu}_{n}^{(-)}\\
			n \ \mathsf{\mu}_{n+1}^{(-)}
		\end{bmatrix},
	\end{equation}
	where $\mathsf{\mu}_{n}^{(\pm)} \equiv \langle\!\langle \Phi^{n\pm 1}\rangle\!\rangle / \langle\!\langle \Phi^{n}\rangle\!\rangle$ is the ratio of two cumulants of consequtive orders. Importanty, from this expression, one can extract the leading partition function zeros from measurements (or simulations) of four high cumulants of the thermodynamic observable $\Phi$.
	
	In the following, we extract the leading partition function zeros from the fluctuations of the energy and the magnetization in the Curie-Weiss model at finite size. By doing so with increasing system size, we can determine the convergence point in the thermodynamic limit. However, to do so, we need a scaling analysis of the partition function zeros and their approach to the real-axis.
	
	\begin{figure*}[t]
		\centering
		\includegraphics[width=1\linewidth]{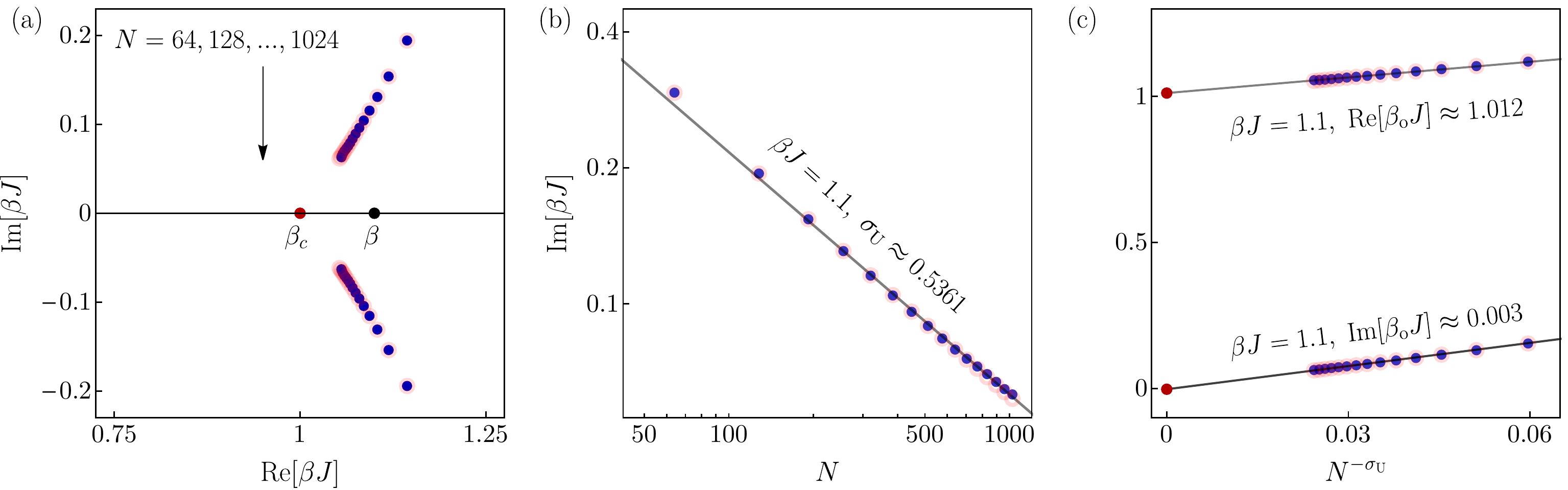}
		\caption{Fisher zeros and critical exponents. (a) Leading Fisher zeros (blue circles) in the complex plane of the inverse temperature with increasing system size $N=64,128,...,1024$. The Fisher zeros are obtained from the energy cumulants of order $n=11,12,13,14$ at the inverse temperature $\beta J=1.1$ (black circle). The convergence point $\beta_c$ in the thermodynamic limit is indicated with a red circle. We also show numerically exact results for the Fisher zeros (red circles), which lie on top of the extracted Fisher zeros. (b) From the finite-size scaling of the imaginary part of the Fisher zeros, we obtain the critical exponent $\sigma_{\rm U}\simeq 0.5361$, which is close to the exact value, $\sigma_{\rm U} =1/2$. (c) Having determined the critical exponent, we can extrapolate the convergence point of the Fisher zeros in the thermodynamic limit, which is close to the exact result $\mathrm{Re} [\beta_c J]=1$.}
		\label{fig:fig3}
	\end{figure*}
	
	\section{FINITE-SIZE SCALING}
	\label{sec:scaling}
	
	\revision{To develop our scaling analysis, we start with the Ising model in $d$ dimensions, before employing the mean-field approximation. For a rectangular lattice of linear size $L$, the number of sites is $N=L^d$. The Privman-Fisher scaling ansatz for the singular part of the free energy per site and the correlation length then reads \cite{Kadanoff1966, Domb1983, Privman1984, Privman1990, Cardy1996, Brankov2000},
		\begin{equation}
				f(t,h,L)=L^{-d}\ \tilde{Y}\left(c_1 t L^{1/\nu},c_2 h L^{\Delta/\nu}\right)
		\end{equation}	
				and
				\begin{equation}
				\xi(t,h,L)=L\ \tilde{X}\left(c_1 t L^{1/\nu},c_2 h L^{\Delta/\nu}\right),
		\end{equation}
		where $\Delta= \mathcal{B} + \gamma$ is expressed in terms of the universal critical exponents, $\mathcal{B}$ and $\gamma$, that characterize the scaling of the magnetization and the susceptibility, respectively,  the critical exponent related to the correlation length is denoted by $\nu$, and $c_{1,2}$ are system dependent non-universal parameters. Above, we have absorbed the factors of $\beta$ into the free energy and the magnetic field by redefining them as $f\rightarrow \beta f$ and $h\rightarrow \beta h$, and we have introduced the reduced temperature, $t=|T-T_c|/T_c$, where $T_c$ is the critical temperature. This ansatz holds for systems below the upper critical dimension, $d<d_c$, where $d_c=4$ for the Ising universality class, and it would not apply for the mean-field approximation of the Ising model. In particular, the hyperscaling relation, $d \nu = 2-\alpha$, where $\alpha$ is the critical exponent of the specific heat, is embedded in this ansatz, but it breaks down for $d>d_c$. To address this issue, Kenna and Berche showed that a modified hyperscaling relation can be formulated as \cite{Kenna2014}
		\begin{equation}
			d\nu = \qoppa (2 - \alpha),
			\label{eq:modhyper}
		\end{equation}	
		where $\qoppa=1$ for $d < d_c$ and $\qoppa=d/d_c$ otherwise. Accordingly, we modify the Privman-Fisher ansatz as \cite{Berche2012, Kenna2014}, 
		\begin{equation}
				f(t,h,L)=L^{-d}\ \widetilde{Y}\left(c_1 t L^{\qoppa/\nu}, c_2 h L^{\qoppa \Delta/\nu}\right)
				\end{equation}
				and
				\begin{equation}
				\xi(t,h,L)=L^\qoppa \widetilde{X}\left(c_1 t L^{\qoppa/\nu},c_2 h L^{\qoppa \Delta/\nu}\right).
		\end{equation}
		From this ansatz, we can obtain the critical behavior of thermodynamic observables of the mean-field models and hypercubic lattices with $d>4$. For instance, with $h=t=0$, the singular part of the specific heat is given as
		\begin{equation}
			c_v \propto \partial^2_t f(t,0,L)|_{t=0} \propto L^{-d} L^{2\qoppa /\nu} \propto L^{\qoppa \alpha/\nu},
		\end{equation}
		while the magnetic susceptibility reads
		\begin{equation}
			\chi \propto \partial^2_h f(0,h,L)|_{h=0} \propto L^{-d} L^{2\qoppa \Delta/\nu} \propto L^{\qoppa \gamma/\nu},
		\end{equation}
		having used the two modified hyperscaling relations, Eq.~(\ref{eq:modhyper}) and  $d\nu=\qoppa(2\mathcal{B}+\gamma)=\qoppa (2\Delta-\gamma)$.
		
		Next, we consider the finite-size scaling of the energy fluctuations close to the critical point. To this end, we make an ansatz for the probability distribution of the singular part of the total energy $U_s$, following Binder~\cite{Binder1981},
		\begin{equation}
			P(U_s, L)=a L^{x} \tilde{p}(b L^{x} U_s, L^\qoppa / \xi)
		\end{equation}
		where $x$ is the scaling exponent, the scaling function is denoted by $\tilde{p}$ , and $a$ and $b$ are constants. The moments of the singular part of the energy follow from the probability distribution as
		\begin{equation}
			\label{eq:defMoments}
			\left\langle U_s^{n}\right\rangle=\int dU_s~ U_s^{n} P(U_s, L),
		\end{equation}
		from which we obtain scaling relations of the form
		\begin{equation}
			\langle U_s^n \rangle = L^{-nx}  v_n(L^\qoppa/\xi).
		\end{equation}
		Using the relation between moments and cumulants,
		\begin{equation}
			\langle\!\langle U^n \rangle\!\rangle=\langle U^{n} \rangle-\sum_{m=1}^{n-1}\left(\begin{array}{l}{n-1} \\ {m-1}\end{array}\right) \langle\!\langle U^m \rangle\!\rangle \langle U^{n-m} \rangle,
		\end{equation}
		we obtain a similar scaling behavior for the cumulants
		\begin{equation}
			\label{eq:scalingU0}
			\langle\!\langle U_s^n \rangle\!\rangle = L^{-nx} ~ u_n(L^\qoppa/\xi),
		\end{equation}
		where the factor $u_n$ is important to obtain  the correct scaling behavior in the thermodynamic limit as we show below. (We note that the cumulants may also have a contribution from the non-singular part of the free energy. However, that will not be important for the high cumulant orders that we consider.) We now proceed by considering the singular part of the specific heat capacity in the thermodynamic limit,
		\begin{equation}
			c_V=\frac{k_B \beta^2}{N}  \langle\!\langle U_s^2\rangle\!\rangle \propto \xi^{\alpha/\nu}.
		\end{equation}
		 Next, we use that $\langle\!\langle U_s^2 \rangle\!\rangle = L^{-2x} u_2(L^\qoppa/\xi)\propto L^d \xi^{\alpha/\nu}$, which implies that $u_2(L^\qoppa/\xi) \propto (L^\qoppa/\xi)^{-\alpha/\nu}$, since both sides should scale in the same way with $\xi$. Consequently, we find the relation $-2x=d+\qoppa \alpha/\nu$ between the scaling exponents. Furthermore, employing the modified hyperscaling relation, we obtain that $x=-\qoppa/\nu$ and thus $\langle\!\langle U^n \rangle\!\rangle =L^{n \qoppa/\nu} u_n(L/\xi)$. Since $\qoppa=d/4$, we finally arrive at the scaling relation
		\begin{equation}
			\label{eq:scalingU}
			\langle\!\langle U_s^n \rangle\!\rangle \propto N^{n \sigma_{\rm U}},
		\end{equation}
		where $\sigma_{\rm U}=1/4\nu$ is a critical exponent that describes the scaling of the energy cumulants. Moreover, since $\nu=1/2$ for mean-field models, we expect to find $\sigma_{\rm U}=1/2$. 
		
		For high cumulant orders, we expect the cumulants to be determined by the singular parts given by Eq.~(\ref{eq:scalingU}). Thus, comparing this scaling relation with the expression in Eq.~(\ref{eq:qCumulants}), we expect that the leading Fisher zeros should obey the scaling relations
		\begin{equation}
			\begin{split}
				|\beta_{\rm o} - \beta_c| &\propto N^{-\sigma_{\rm U}}, \\
				\rm{Im}\left[\beta_{\rm o}\right]  &\propto N^{-\sigma_{\rm U}}.
			\end{split}
		\end{equation}
		Thus, with these relations, we can extract the critical exponent $\sigma_{\rm U}$ from the size dependence of the leading Fisher zeros. Subsequently, we can use the same relations to accurately extrapolate the position of the leading Fisher zeros in the thermodynamic limit and thereby determine the temperature at which a phase transition will occur.
		
		For the magnetization, we again make the ansatz 
		\begin{equation}
			P(M, L)=aL^{x}  \tilde{p}(b L^{x}M, L^\qoppa/ \xi).
		\end{equation} 
		Using a similar argument as above, we obtain scaling relations for the magnetization cumulants
		\begin{equation}
			\label{eq:scalingM}
			\langle\!\langle M^n \rangle\!\rangle  \propto N^{n \sigma_{\rm M}},
		\end{equation}
		having expressed $x=-\qoppa \Delta /\nu$ in terms of the critical exponent of the magnetization $\delta$ and defined $\sigma_{\rm M}=\delta/(1+\delta)$ \cite{Krasnytska2016}. For the Lee-Yang zeros, we then find
		\begin{equation}
			\begin{split}
				|h_{\rm o} - h_c| &\propto N^{-\sigma_{\rm M}}, \\
				\mathrm{Im}\left[ h_{\rm o}\right]  &\propto N^{-\sigma_{\rm M}},
			\end{split}
		\end{equation}
		which we can use to determine the critical exponent $\sigma_{\rm M}$ from the Lee-Yang zeros as well as their convergence point in the thermodynamic limit. Since $\delta=3$ for mean-field models, we expect to find $\sigma_{\rm M}=3/4$. We note that the same relation for the Lee-Yang zeros of the Curie-Weiss model was reported in Ref.~\cite{Krasnytska2016}.
	}
	
	\section{FISHER ZEROS \& CRITICAL EXPONENTS}
	\label{sec:fisher}

\begin{figure*}
	\centering
	\includegraphics[width=1\linewidth]{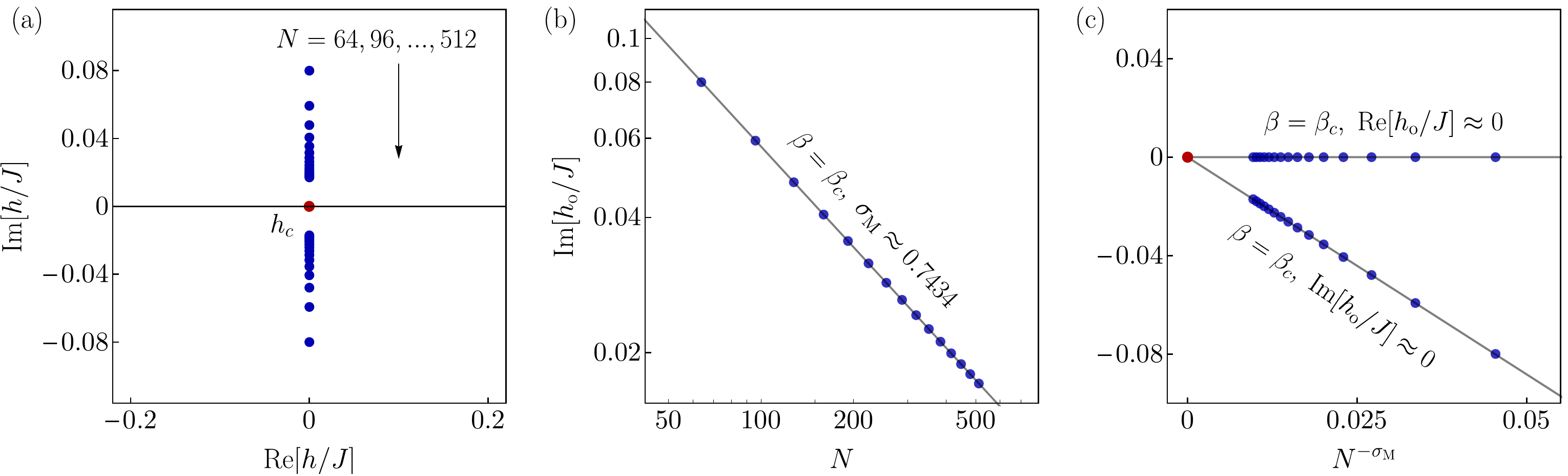}
	\caption{Lee-Yang zeros and critical exponents. (a) Leading Lee-Yang zeros (blue circles) in the complex plane of the magnetic field with increasing system size $N=64,96,...,512$. The Lee-Yang zeros are extracted from magnetization cumulants of order $n=11,12,13,14$ with the external field $h/J=0.002$ and the inverse temperature $\beta=\beta_c$. The convergence point in the thermodynamic limit is indicated with a red circle, $h_c$. (b) From the finite-size scaling of the imaginary part of the Lee-Yang zeros (blue circles), we obtain the critical exponent $\sigma_{\rm M}\simeq 0.7434$, which is close to the exact value of $\sigma_{\rm M} =3/4$. (c) Both real and imaginary parts of the Lee-Yang zeros vanish in the thermodynamic limit, signaling a phase transition at $h_c=0$.}
	\label{fig:fig5}
\end{figure*}
	
	We are now ready to determine the Fisher zeros for the Curie-Weiss model from the energy fluctuations in the system at finite size. We first note that the cumulants are obtained as logarithmic derivatives of the partition function with respect to $q=-\beta$, such that 
	\begin{equation}\label{eq:engCumulants}
		\langle\!\langle U^n\rangle\!\rangle =\partial_{-\beta}^n \ln Z.
	\end{equation}
	In the following, we evaluate the high cumulants numerically using the partition function for finite system-sizes. We then extract the leading Fisher zeros from the energy fluctuations using Eq.~(\ref{eq:QMethod}), which in this case reads
	\begin{equation}
		\label{eq:QMethodEnergy}
		\begin{bmatrix}
			2~\mathrm{Re}[\beta-\beta_{\rm o}]\\
			|\beta-\beta_{\rm o}|^2
		\end{bmatrix}=\begin{bmatrix}
			1& -\frac{\mathsf{\kappa}_{n}^{(+)}}{n}\\
			1& -\frac{\mathsf{\kappa}_{n+1}^{(+)}}{n+1}
		\end{bmatrix}^{-1}
		\begin{bmatrix}
			(n-1) \mathsf{\kappa}_{n}^{(-)}\\
			n \ \mathsf{\kappa}_{n+1}^{(-)},
		\end{bmatrix},
	\end{equation}
	where $\mathsf{\kappa}_{n}^{(\pm)} \equiv \langle\!\langle U^{n\pm 1}\rangle\!\rangle / \langle\!\langle U^{n}\rangle\!\rangle$ is the ratio of subsequent energy cumulants. Importantly, this relation makes it possible to extract the Fisher zeros from cumulants that have been obtained (in an experiment or from simulations) at a single fixed temperature, which can be below or above the critical temperature. As such, our method may potentially enable experimental investigations of phase transitions that are hard to reach, for instance at very low temperatures.

	In Fig.~\ref{fig:fig3}(a), we show the leading Fisher zeros in the complex plane of the inverse temperature, extracted from the high cumulants of the energy. With increasing system size, the zeros approach the real line, which they eventually reach in the thermodynamic limit. We compare the extracted Fisher zeros with numerically exact results and find very good agreement. However, it should be stressed that our method does not rely on an explicit expression for the partition function, but it can be directly applied to cumulants of the energy fluctuations, which can be experimentally measured. To extrapolate the exact convergence point on the real-axis, we proceed in Fig.~\ref{fig:fig3}(b) with a scaling analysis of the imaginary part of the Fisher zeros. From this analysis we obtain the critical exponent~$\sigma_{\rm U}$, which comes close to the exact value of $\sigma_{\rm U}=1/2$, since it is known that $\alpha=0$ for the Curie-Weiss model~\cite{friedli_velenik_2017}. Finally, in Fig.~\ref{fig:fig3}(c), we use the scaling relations for the Fisher zeros to determine the convergence points in the thermodynamic limit. The imaginary part essentially vanishes in the thermodynamic limit, while the real part comes very close to the exact critical value of $\beta_c J = 1$. We stress that these results are obtained at a single fixed temperature below the critical temperature, and unlike the conventional use of Binder cumulants~\cite{Binder1981,Binder1984,Binder2001}, we do not need to tune the temperature across the critical value to extract the critical exponents. \revision{In Fig.~\ref{fig:fig6} near the end of the paper, we discuss error estimates, including the influence of system size, temperature, and cumulant orders.}
	
	\section{LEE-YANG ZEROS \& CRITICALITY}
	\label{sec:lee-yang}
	
	Next, we turn to the Lee-Yang zeros in the complex plane of the magnetic field. In this case, we extract the Lee-Yang zeros from the cumulants of the magnetization, which are obtained as logarithmic derivatives of the partition function with respect to $q=\beta h$ as
	\begin{equation}
		\label{eq:magCumulants}
		\langle\!\langle M^n\rangle\!\rangle =\partial_{h}^n \ln Z/\beta^n.
	\end{equation}
	We can then extract the leading Lee-Yang zeros from the magnetization cumulants using the cumulant method given by Eq.~(\ref{eq:QMethod}), which takes the explicit form
	\begin{equation}
		\label{eq:QMethodMag}
		\begin{bmatrix}
			2\beta\mathrm{Re}[ h_{\rm o}- h]\\
			\beta^2| h_{\rm o}- h|^2
		\end{bmatrix}=\begin{bmatrix}
			1& -\frac{\mathsf{\omega}_{n}^{(+)}}{n}\\
			1& -\frac{\mathsf{\omega}_{n+1}^{(+)}}{n+1}
		\end{bmatrix}^{-1}
		\begin{bmatrix}
			(n-1) \mathsf{\omega}_{n}^{(-)}\\
			n \ \mathsf{\omega}_{n+1}^{(-)}
		\end{bmatrix},
	\end{equation}
	where $\mathsf{\omega}_{n}^{(\pm)} \equiv \langle\!\langle M^{n\pm 1}\rangle\!\rangle / \langle\!\langle M^{n}\rangle\!\rangle$ again is the ratio of two cumulants of subsequent orders.
	
	In Fig.~\ref{fig:fig5}(a), we show the extracted Lee-Yang zeros in the complex plane of the magnetic field. The system is at the critical temperature $\beta=\beta_c$ and the zeros approach the critical value at $h_c=0$ on a straight line that is perpendicular to the real-axis, signaling a first-order phase transition as a function of the magnetic field~\cite{Huang1987,Biskup2000,Janke2001}. Again, we perform a scaling analysis of the imaginary part of the Lee-Yang zeros in Fig.~\ref{fig:fig5}(b) to determine the critical exponent $\sigma_{\rm M}$, which comes close to the exact value of $\sigma_{\rm M}=3/4$, since it is known that $\delta=3$ for the Curie-Weiss model~\cite{friedli_velenik_2017}. With the critical exponent at hand, we can finally extrapolate the position of the leading Lee-Yang zeros in the thermodynamic limit and we indeed find that they converge to the value of $h_c\simeq 0$.
	
	In Fig.~\ref{fig:fig4}, we show the Lee-Yang zeros above, below, and at the critical temperature. At low temperatures, the Lee-Yang zeros converge to the real axis, corresponding to a first-order phase transition at $h=0$. Above the critical temperature, the Lee-Yang zeros remain complex in the thermodynamic limit, since there is no phase transition. 
	
	\section{LARGE-DEVIATION STATISTICS}
	\label{sec:LDF}
	
\begin{figure}
		\centering
		\includegraphics[width=0.9\columnwidth]{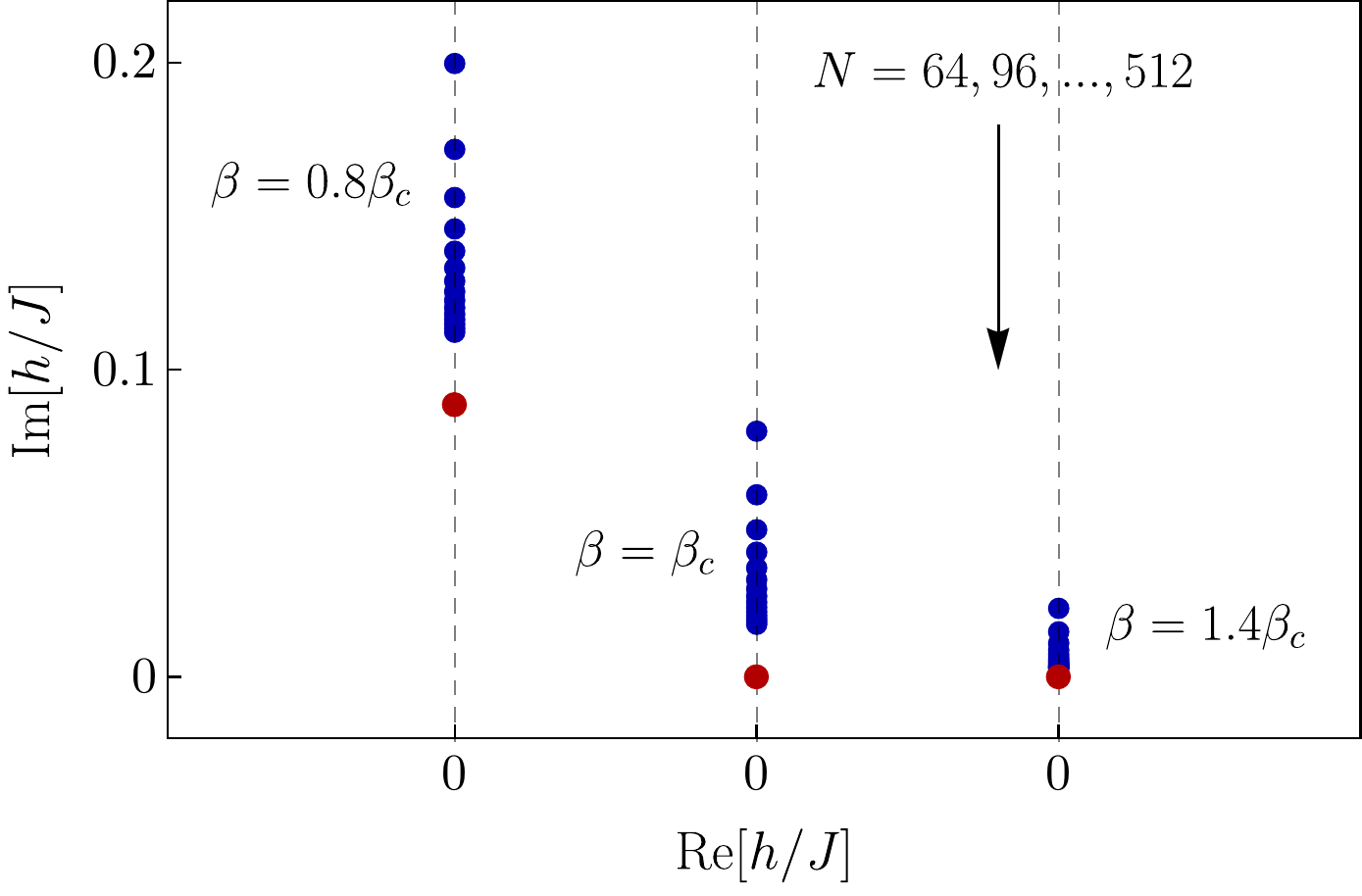}
		\caption{Lee-Yang zeros above, at, and below the critical temperature. For the sake of clarity, the results for different temperatures have been shifted horizontally as indicated by dashed lines. Above the critical temperature, the zeros remain complex, and there is no phase transition. By contrast, below and at the critical temperature, the Lee-Yang zeros reach $h_c\simeq0$, corresponding to a sharp phase transition.}
		\label{fig:fig4}
	\end{figure}

	\revision{Having determined the partition function zeros for the Curie-Weiss model, we will now see how they relate to the large-deviation statistics of the magnetization. Recently, we showed that the large-deviation statistics of the energy fluctuations close to a first-order phase transition are encoded in the convergence points of the Fisher zeros~\cite{Deger2018}. We now investigate, if similar relations hold between the rare fluctuations of the magnetization and the Lee-Yang zeros. Such relations would suggest that a profound connection between Lee-Yang theory and large-deviation statistics may exist. To proceed, we first write} the probability density for the magnetization as
	\begin{equation}
		P(M)=\sum_{\{\sigma_i\}} \frac{e^{-\beta U(\{\sigma_i\})}}{Z(\beta,h)}\int_{-\pi}^\pi \frac{d\chi}{2\pi} e^{i\chi\left(\sum_{i}\sigma_i-M\right)},
	\end{equation}
	where we have used an integral form of the Kronecker delta. Shifting the argument of the partition function as $h\rightarrow h+i\chi/\beta$, we can write the distribution as~\cite{Deger2018}
	\begin{equation}
		P(M)=\frac{1}{2 \pi} \int_{-\pi}^\pi d\chi ~ \frac{Z\left(\beta,h+i\chi/ \beta\right)}{Z\left(\beta,h\right)}  e^{-i \chi M}.
	\end{equation}
	having used that
	\begin{equation}
		Z\left(\beta,h+i \chi/\beta\right) = \sum_{\{\sigma_i\}} e^{ -\beta U\left( \{\sigma_i\} \right) } e^{i \chi \sum_i \sigma_i}.
	\end{equation}
	Using the substitution $\kappa= h+i\chi/\beta$, we obtain an expression for the probability distribution reading
	\begin{equation}
		P(M)=\frac{\beta}{2 \pi i} \int_{h-i \pi / \beta}^{h+i \pi / \beta} d\kappa ~ \frac{Z\left(\beta,\kappa\right)}{Z\left(\beta,h\right)}  e^{M \beta \left(h-\kappa\right)}.
	\end{equation}
	Furthermore, writing the partition function in terms of the free energy as $\ln Z=-\beta F$, we find the expression
	\begin{equation}
		P(m) = \frac{\beta}{2 \pi i} \int_{h-i \pi / \beta}^{h+i \pi / \beta} d\kappa ~ e^{N \Theta(\kappa,m)},
	\end{equation}
	having defined the exponent of the integrand as
	\begin{equation}
		\label{eq:LDF-theta}
		\Theta(\kappa,m)=\beta \left[ f(h) + h m \right]-\beta \left[ f(\kappa) + \kappa m \right],
	\end{equation}
	where $f\equiv F/N$ and $m \equiv M/N$ are the free energy and the magnetization per site. In the thermodynamic limit, we can now evaluate the large deviation function as
	\begin{equation}
		\frac{\ln P(m)}{N}\simeq\Theta(\kappa_0,m)
	\end{equation}
	where $\kappa_0=\kappa_0(m)$ solves the saddle-point equation $\partial_\kappa \Theta(\kappa,m)=0$. In general, it is challenging to solve the saddle-point equation. However, we can follow our recent work on Lee-Yang theory and make the ansatz~\cite{Deger2018}
	\begin{equation}
		\Theta(\kappa,m)\simeq \mathcal{M}+m \beta (h-\kappa)-\bar{m}  \beta\sqrt{(h_{c}-\kappa ) \left(h_{c}^*-\kappa \right)}
		\label{eq:ansatz}
	\end{equation}
	where $h_{c}$ and $h_{c}^{*}$ are the convergence points of the Lee-Yang zeros in the thermodynamic limit, and $\mathcal{M}$ and $\bar{m}$ are (unknown) constants. Here, the expectation is that the free energy in the thermodynamic limit has square-root branch points at the convergence points of the Lee-Yang zeros, for instance, as in the case of an eigenvalue crossing of a transfer matrix. Under these assumptions, we can solve the saddle-point equation, and we then find an expression for the large-deviation statistics reading
	\begin{equation}
		\frac{\ln P(m)}{N}=\mathcal{M}+ m\beta \left[h-\mathrm{Re}(h_c)\right] - \beta|\mathrm{Im}(h_c)| \frac{\bar{m}^2 + m^2}{\sqrt{\bar{m}^2-m^2}}
		\label{eq:LDFapprox}
	\end{equation}
	in terms of the Lee-Yang zeros and the applied magnetic field. Thus, for the three different temperatures used in Fig.~\ref{fig:fig4}, we can insert the extracted convergence points and adjust the unknown parameters so that the analytic expression matches the numerically exact results.
	
	\begin{figure*}
		\centering
		\includegraphics[width=0.95\textwidth]{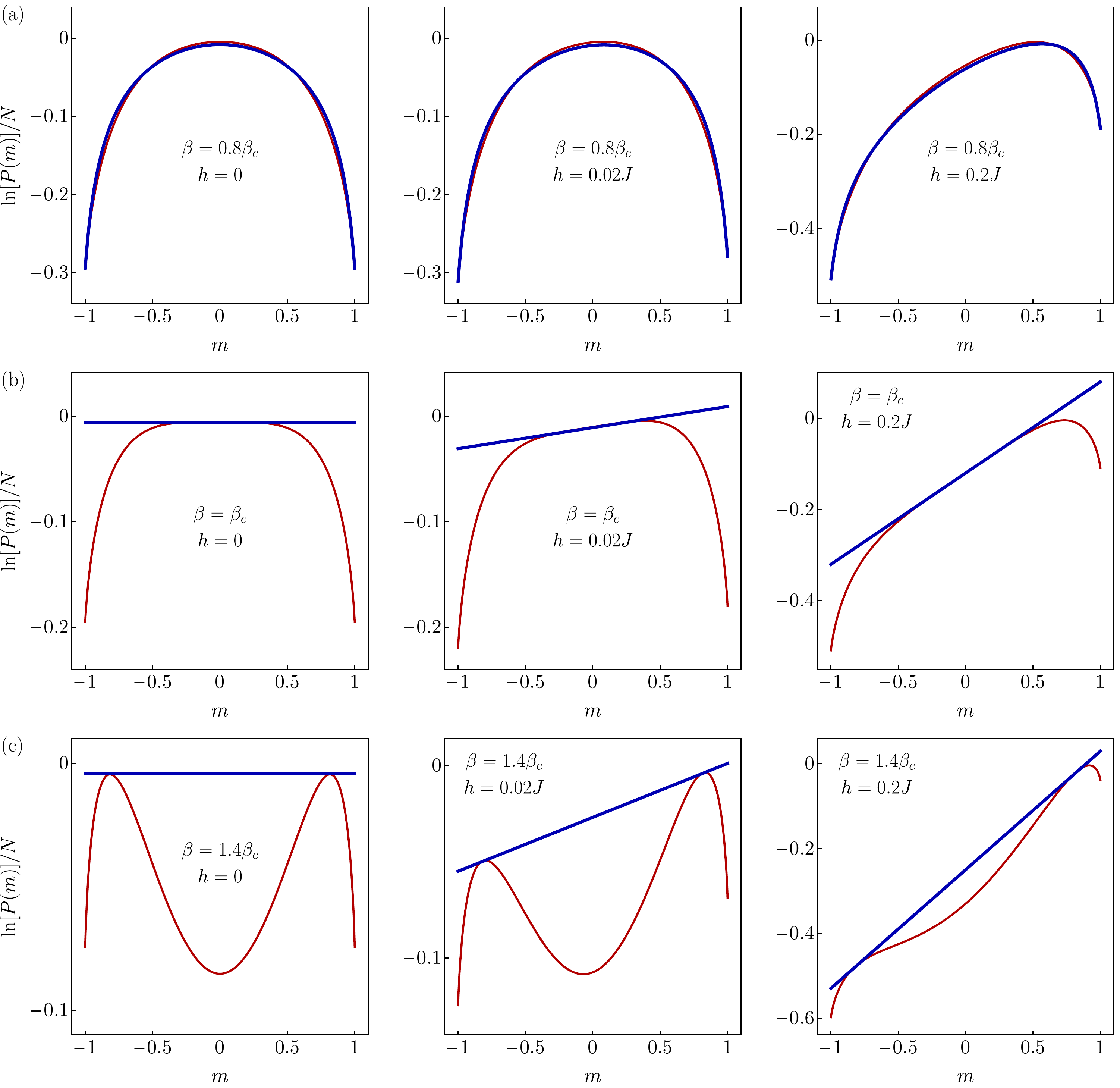}
		\caption{Large-deviation statistics of the magnetization. Numerically exact results are shown in red (light gray), while the approximation in Eq.~(\ref{eq:LDFapprox}) is shown in blue (dark gray), using the convergence points extracted in Fig.~\ref{fig:fig6}. In row (a) we have used $\bar{m}=1.085$ to fit the tails of the distribution, while $\mathcal{M}$ is used to shift the curves vertically. The temperature (from top to bottom) is above, at, and below the critical temperature. The magnetic field increases from left to right. 
		}
		\label{fig:fig7}
	\end{figure*}
	
	Figure \ref{fig:fig7} shows the results of this procedure. Despite being a crude approximation, Eq.~(\ref{eq:LDFapprox}) captures many essential features of the large-deviation statistics. In Fig.~\ref{fig:fig7}(a), the temperature is above the critical temperature, and the Lee-Yang zeros remain complex in the thermodynamic limit. In the left panel, we first fix the parameter $\bar{m}$, which controls the tails of the distribution. (The vertical shift of the curves is controlled by $\mathcal{M}$, which is adjusted in each panel.) Having fixed this parameter, we can apply a magnetic field, and we then see how the analytical expression nicely captures the exact results in the middle and right panels with an increasing magnetic field.  
	
	As the temperature is lowered, the Lee-Yang zeros eventually reach the real-axis, and the large-deviation statistics develop a nearly flat plateau. In Fig.~\ref{fig:fig7}(b), we show exact results for the large-deviation statistics at the critical temperature together with the straight line predicted by our analytical approximation. In this case, the approximation captures the flat plateau of the distribution. On the other hand, it does not describe the tails of the distribution, which are governed by the fluctuations around the spin configurations with a positive or a negative average magnetization. As the temperature is further decreased, the Lee-Yang zeros remain real, however, the distribution of the magnetization now becomes bimodal, which is not accounted for by our approximation, which only captures  the convex hull of the large-deviation statistics. Technically, we evaluate the large-deviation statistics using a Legendre transformation, which can only produce an upper-convex function (with our sign convention) \cite{Touchette2009}. On the other hand, it is well-known that systems with long-range interactions, including mean-field models, may have non-concave entropies and, as a result, the large-deviation statistics can be bimodal \cite{Touchette_2008,Touchette_2010}. For the Curie-Weiss model, all spins interact however far apart, and as a result, the large-deviation function becomes bimodal below the critical temperature. We expect that for spin lattices with short-range interactions, for example the Ising model~\cite{Alves2000,GarciaSaez2015,Xu2019,Goldenfeld2018}, the large-deviation function will be upper-convex also below the critical temperature. 

\begin{figure*}
	\centering
	\includegraphics[width=1\textwidth]{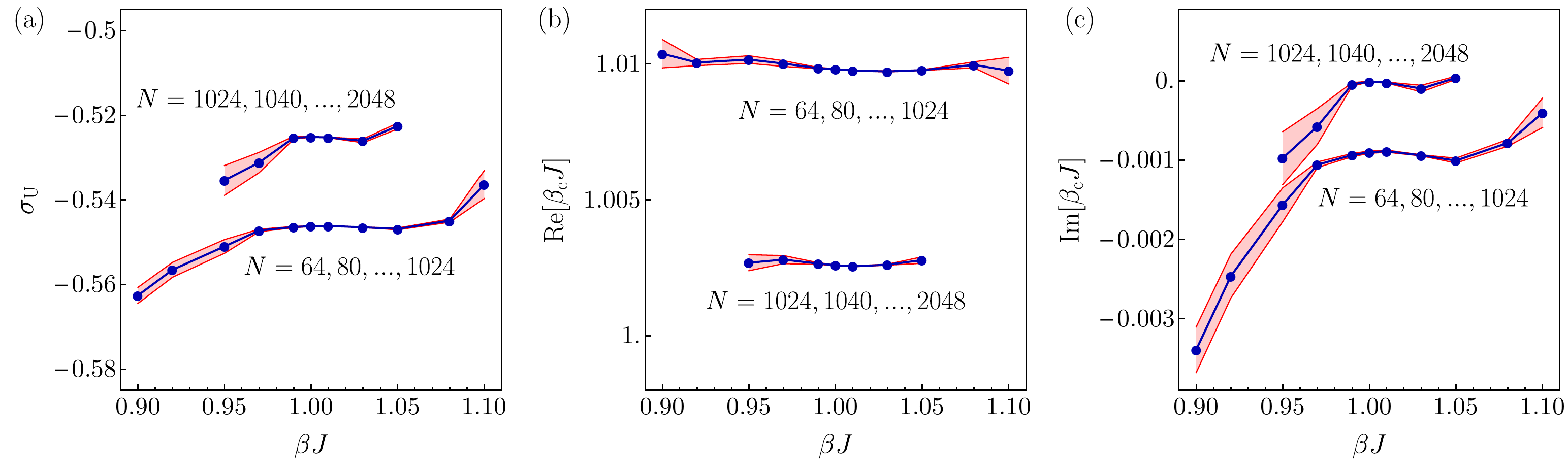}
	\caption{\revision{Error analysis for the Fisher zeros with different system sizes, temperatures, and cumulant orders. The red bands indicate the standard error, $\sigma/\sqrt{m}$, where $\sigma$ denotes the standard deviation and $m=7$ is the number of different sets of high-order cumulants used to extract the Fisher zeros, i.e., $n=11,\ldots 14;12,\ldots, 15;...;17,\ldots, 20$. The mean values over the cumulant sets are indicated by blue points at the inverse temperatures $\beta J=0.9,0.92,0.95,0.97,0.99,1,1.01,1.03,1.05,1.08,1.1$. Each panel consists of two different sets of system sizes in order to investigate finite-size effects. (a) The critical exponent $\sigma_{\rm U}$ is extracted from log-log plots of the imaginary part of the Fisher zeros as a function of the system size. (b) The critical point is estimated to be ${\rm Re} [\beta_{\rm c} J] \approx 1.00976$ by averaging over the small system sizes and ${\rm Re} [\beta_{\rm c} J] \approx 1.00260$ for the large system sizes. (c) The convergence point of the imaginary part of the Fisher zeros yields ${\rm Im} [\beta_{\rm c} J] \approx 0$ to a good approximation for both sizes.}}
	\label{fig:fig6}
\end{figure*}
	
	\section{CONCLUSIONS}
	\label{sec:conclusions}
	
	We have investigated the Curie-Weiss model of spontaneous magnetization using our recently established cumulant method. In particular, we have shown how the Fisher zeros and the Lee-Yang zeros of the Curie-Weiss model can be extracted from the fluctuations of the energy and the magnetization, respectively, in systems of finite size. Importantly, since these observables are measurable, our method provides a direct link from experiments (or simulations) to the determination of partition function zeros of critical systems. Based on a finite-size scaling analysis, we have determined both the critical point and the critical exponents of the Curie-Weiss model from the approach of the partition function zeros to the real-axis. Our method can be employed with fixed control parameters, and there is no need to tune the system across the phase transition. For this reason, our method may enable experimental investigations of phase transitions that may be hard to reach, for instance at very low temperatures or very high pressures. Finally, we have shown that the Lee-Yang zeros carry important information about the rare fluctuations of the magnetization. Specifically, the convergence points of the Lee-Yang zeros in the thermodynamic limit capture many essential features of the large-deviation statistics of the magnetization, including its dependence on the magnetic field. \revision{This finding suggests that a profound connection between Lee-Yang theory and large-deviation statistics may exist.}
	
	Our work is an important test of the cumulant method, and it illustrates that the method indeed has a broad scope of potential applications. In future work, it would be interesting to apply the method to dynamical phase transitions in quantum many-body systems after a quench~\cite{Heyl2013,Zvyagin2016,Heyl2018} and to quantum phase transitions in the groundstate of interacting quantum systems~\cite{Lamacraft2008}. 
	
	\begin{acknowledgments}
		We thank A. Acharya, K.~Brandner, F.~Brange, J.~P.~Garrahan, and R.~Kenna for insightful discussions. We acknowledge the computational resources provided by the Aalto Science-IT project. A.~D.~and C.~F.~are associated with Centre for Quantum Engineering at Aalto University. A.~D.~acknowledges support from the Vilho, Yrjö and Kalle Väisälä Foundation of the Finnish Academy of Science and Letters through the grant for doctoral studies. The work was supported by the Academy of Finland (projects No.~308515 and 312299).
	\end{acknowledgments}


\begin{thebibliography}{60}%
	\makeatletter
	\providecommand \@ifxundefined [1]{%
		\@ifx{#1\undefined}
	}%
	\providecommand \@ifnum [1]{%
		\ifnum #1\expandafter \@firstoftwo
		\else \expandafter \@secondoftwo
		\fi
	}%
	\providecommand \@ifx [1]{%
		\ifx #1\expandafter \@firstoftwo
		\else \expandafter \@secondoftwo
		\fi
	}%
	\providecommand \natexlab [1]{#1}%
	\providecommand \enquote  [1]{``#1''}%
	\providecommand \bibnamefont  [1]{#1}%
	\providecommand \bibfnamefont [1]{#1}%
	\providecommand \citenamefont [1]{#1}%
	\providecommand \href@noop [0]{\@secondoftwo}%
	\providecommand \href [0]{\begingroup \@sanitize@url \@href}%
	\providecommand \@href[1]{\@@startlink{#1}\@@href}%
	\providecommand \@@href[1]{\endgroup#1\@@endlink}%
	\providecommand \@sanitize@url [0]{\catcode `\\12\catcode `\$12\catcode
		`\&12\catcode `\#12\catcode `\^12\catcode `\_12\catcode `\%12\relax}%
	\providecommand \@@startlink[1]{}%
	\providecommand \@@endlink[0]{}%
	\providecommand \url  [0]{\begingroup\@sanitize@url \@url }%
	\providecommand \@url [1]{\endgroup\@href {#1}{\urlprefix }}%
	\providecommand \urlprefix  [0]{URL }%
	\providecommand \Eprint [0]{\href }%
	\providecommand \doibase [0]{http://dx.doi.org/}%
	\providecommand \selectlanguage [0]{\@gobble}%
	\providecommand \bibinfo  [0]{\@secondoftwo}%
	\providecommand \bibfield  [0]{\@secondoftwo}%
	\providecommand \translation [1]{[#1]}%
	\providecommand \BibitemOpen [0]{}%
	\providecommand \bibitemStop [0]{}%
	\providecommand \bibitemNoStop [0]{.\EOS\space}%
	\providecommand \EOS [0]{\spacefactor3000\relax}%
	\providecommand \BibitemShut  [1]{\csname bibitem#1\endcsname}%
	\let\auto@bib@innerbib\@empty
	%</preamble>
	\bibitem [{\citenamefont {Lee}\ and\ \citenamefont {Yang}(1952)}]{Lee1952}%
	\BibitemOpen
	\bibfield  {author} {\bibinfo {author} {\bibfnamefont {T.~D.}\ \bibnamefont
			{Lee}}\ and\ \bibinfo {author} {\bibfnamefont {C.~N.}\ \bibnamefont {Yang}},\
	}\bibfield  {title} {\enquote {\bibinfo {title} {Statistical {T}heory of
				{E}quations of {S}tate and {P}hase {T}ransitions. {II}. {L}attice {G}as and
				{I}sing {M}odel},}\ }\href {\doibase 10.1103/PhysRev.87.410} {\bibfield
		{journal} {\bibinfo  {journal} {Phys. Rev.}\ }\textbf {\bibinfo {volume}
			{87}},\ \bibinfo {pages} {410} (\bibinfo {year} {1952})}\BibitemShut
	{NoStop}%
	\bibitem [{\citenamefont {Yang}\ and\ \citenamefont {Lee}(1952)}]{Yang1952a}%
	\BibitemOpen
	\bibfield  {author} {\bibinfo {author} {\bibfnamefont {C.~N.}\ \bibnamefont
			{Yang}}\ and\ \bibinfo {author} {\bibfnamefont {T.~D.}\ \bibnamefont {Lee}},\
	}\bibfield  {title} {\enquote {\bibinfo {title} {Statistical {T}heory of
				{E}quations of {S}tate and {P}hase {T}ransitions. {I}. {T}heory of
				{C}ondensation},}\ }\href {\doibase 10.1103/PhysRev.87.404} {\bibfield
		{journal} {\bibinfo  {journal} {Phys. Rev.}\ }\textbf {\bibinfo {volume}
			{87}},\ \bibinfo {pages} {404} (\bibinfo {year} {1952})}\BibitemShut
	{NoStop}%
	\bibitem [{\citenamefont {Blythe}\ and\ \citenamefont
		{Evans}(2002)}]{Blythe2002}%
	\BibitemOpen
	\bibfield  {author} {\bibinfo {author} {\bibfnamefont {R.~A.}\ \bibnamefont
			{Blythe}}\ and\ \bibinfo {author} {\bibfnamefont {M.~R.}\ \bibnamefont
			{Evans}},\ }\bibfield  {title} {\enquote {\bibinfo {title} {{Lee-Yang Zeros
					and Phase Transitions in Nonequilibrium Steady States}},}\ }\href {\doibase
		10.1103/PhysRevLett.89.080601} {\bibfield  {journal} {\bibinfo  {journal}
			{Phys. Rev. Lett.}\ }\textbf {\bibinfo {volume} {89}},\ \bibinfo {pages}
		{080601} (\bibinfo {year} {2002})}\BibitemShut {NoStop}%
	\bibitem [{\citenamefont {Bena}\ \emph {et~al.}(2005)\citenamefont {Bena},
		\citenamefont {Droz},\ and\ \citenamefont {Lipowski}}]{Bena2005}%
	\BibitemOpen
	\bibfield  {author} {\bibinfo {author} {\bibfnamefont {I.}~\bibnamefont
			{Bena}}, \bibinfo {author} {\bibfnamefont {M.}~\bibnamefont {Droz}}, \ and\
		\bibinfo {author} {\bibfnamefont {A.}~\bibnamefont {Lipowski}},\ }\bibfield
	{title} {\enquote {\bibinfo {title} {Statistical mechanics of equilibrium and
				nonequilibrium phase transitions: The {Y}ang-{L}ee formalism},}\ }\href
	{\doibase 10.1142/S0217979205032759} {\bibfield  {journal} {\bibinfo
			{journal} {Int. J. Mod. Phys. B}\ }\textbf {\bibinfo {volume} {19}},\
		\bibinfo {pages} {4269} (\bibinfo {year} {2005})}\BibitemShut {NoStop}%
	\bibitem [{\citenamefont {Lee}(2013{\natexlab{a}})}]{Lee2013}%
	\BibitemOpen
	\bibfield  {author} {\bibinfo {author} {\bibfnamefont {J.}~\bibnamefont
			{Lee}},\ }\bibfield  {title} {\enquote {\bibinfo {title} {Exact {P}artition
				{F}unction {Z}eros of the {W}ako-{S}ait\^o-{M}u\~noz-{E}aton {P}rotein
				{M}odel},}\ }\href {\doibase 10.1103/PhysRevLett.110.248101} {\bibfield
		{journal} {\bibinfo  {journal} {Phys. Rev. Lett.}\ }\textbf {\bibinfo
			{volume} {110}},\ \bibinfo {pages} {248101} (\bibinfo {year}
		{2013}{\natexlab{a}})}\BibitemShut {NoStop}%
	\bibitem [{\citenamefont {Lee}(2013{\natexlab{b}})}]{Lee2013b}%
	\BibitemOpen
	\bibfield  {author} {\bibinfo {author} {\bibfnamefont {J.}~\bibnamefont
			{Lee}},\ }\bibfield  {title} {\enquote {\bibinfo {title} {Exact partition
				function zeros of the wako-sait\^o-mu\~noz-eaton $\ensuremath{\beta}$ hairpin
				model},}\ }\href {\doibase 10.1103/PhysRevE.88.022710} {\bibfield  {journal}
		{\bibinfo  {journal} {Phys. Rev. E}\ }\textbf {\bibinfo {volume} {88}},\
		\bibinfo {pages} {022710} (\bibinfo {year} {2013}{\natexlab{b}})}\BibitemShut
	{NoStop}%
	\bibitem [{\citenamefont {Deger}\ \emph {et~al.}(2018)\citenamefont {Deger},
		\citenamefont {Brandner},\ and\ \citenamefont {Flindt}}]{Deger2018}%
	\BibitemOpen
	\bibfield  {author} {\bibinfo {author} {\bibfnamefont {A.}~\bibnamefont
			{Deger}}, \bibinfo {author} {\bibfnamefont {K.}~\bibnamefont {Brandner}}, \
		and\ \bibinfo {author} {\bibfnamefont {C.}~\bibnamefont {Flindt}},\
	}\bibfield  {title} {\enquote {\bibinfo {title} {Lee-{Y}ang zeros and
				large-deviation statistics of a molecular zipper},}\ }\href {\doibase
		10.1103/physreve.97.012115} {\bibfield  {journal} {\bibinfo  {journal} {Phys.
				Rev. E}\ }\textbf {\bibinfo {volume} {97}},\ \bibinfo {pages} {012115}
		(\bibinfo {year} {2018})}\BibitemShut {NoStop}%
	\bibitem [{\citenamefont {Arndt}\ \emph {et~al.}(2001)\citenamefont {Arndt},
		\citenamefont {Dahmen},\ and\ \citenamefont {Hinrichsen}}]{Arndt2001}%
	\BibitemOpen
	\bibfield  {author} {\bibinfo {author} {\bibfnamefont {P.~F.}\ \bibnamefont
			{Arndt}}, \bibinfo {author} {\bibfnamefont {S.~R.}\ \bibnamefont {Dahmen}}, \
		and\ \bibinfo {author} {\bibfnamefont {H.}~\bibnamefont {Hinrichsen}},\
	}\bibfield  {title} {\enquote {\bibinfo {title} {{Directed percolation,
					fractal roots and the Lee-Yang theorem}},}\ }\href {\doibase
		10.1016/S0378-4371(01)00064-4} {\bibfield  {journal} {\bibinfo  {journal}
			{Physica A}\ }\textbf {\bibinfo {volume} {295}},\ \bibinfo {pages} {128}
		(\bibinfo {year} {2001})}\BibitemShut {NoStop}%
	\bibitem [{\citenamefont {Dammer}\ \emph {et~al.}(2002)\citenamefont {Dammer},
		\citenamefont {Dahmen},\ and\ \citenamefont {Hinrichsen}}]{Dammer2002}%
	\BibitemOpen
	\bibfield  {author} {\bibinfo {author} {\bibfnamefont {S.~M.}\ \bibnamefont
			{Dammer}}, \bibinfo {author} {\bibfnamefont {S.~R.}\ \bibnamefont {Dahmen}},
		\ and\ \bibinfo {author} {\bibfnamefont {H.}~\bibnamefont {Hinrichsen}},\
	}\bibfield  {title} {\enquote {\bibinfo {title} {{Y}ang-{L}ee zeros for a
				nonequilibrium phase transition},}\ }\href {\doibase
		10.1088/0305-4470/35/21/303} {\bibfield  {journal} {\bibinfo  {journal} {J.
				Phys. A: Math. Gen.}\ }\textbf {\bibinfo {volume} {35}},\ \bibinfo {pages}
		{4527} (\bibinfo {year} {2002})}\BibitemShut {NoStop}%
	\bibitem [{\citenamefont {Krasnytska}\ \emph {et~al.}(2015)\citenamefont
		{Krasnytska}, \citenamefont {Berche}, \citenamefont {Holovatch},\ and\
		\citenamefont {Kenna}}]{Krasnytska2015}%
	\BibitemOpen
	\bibfield  {author} {\bibinfo {author} {\bibfnamefont {M.}~\bibnamefont
			{Krasnytska}}, \bibinfo {author} {\bibfnamefont {B.}~\bibnamefont {Berche}},
		\bibinfo {author} {\bibfnamefont {Y.}~\bibnamefont {Holovatch}}, \ and\
		\bibinfo {author} {\bibfnamefont {R.}~\bibnamefont {Kenna}},\ }\bibfield
	{title} {\enquote {\bibinfo {title} {Violation of {Lee-Yang} circle theorem
				for {I}sing phase transitions on complex networks},}\ }\href {\doibase
		10.1209/0295-5075/111/60009} {\bibfield  {journal} {\bibinfo  {journal}
			{EPL}\ }\textbf {\bibinfo {volume} {111}},\ \bibinfo {pages} {60009}
		(\bibinfo {year} {2015})}\BibitemShut {NoStop}%
	\bibitem [{\citenamefont {Krasnytska}\ \emph {et~al.}(2016)\citenamefont
		{Krasnytska}, \citenamefont {Berche}, \citenamefont {Holovatch},\ and\
		\citenamefont {Kenna}}]{Krasnytska2016}%
	\BibitemOpen
	\bibfield  {author} {\bibinfo {author} {\bibfnamefont {M.}~\bibnamefont
			{Krasnytska}}, \bibinfo {author} {\bibfnamefont {B.}~\bibnamefont {Berche}},
		\bibinfo {author} {\bibfnamefont {Y.}~\bibnamefont {Holovatch}}, \ and\
		\bibinfo {author} {\bibfnamefont {R.}~\bibnamefont {Kenna}},\ }\bibfield
	{title} {\enquote {\bibinfo {title} {Partition function zeros for the {I}sing
				model on complete graphs and on annealed scale-free networks},}\ }\href
	{\doibase 10.1088/1751-8113/49/13/135001} {\bibfield  {journal} {\bibinfo
			{journal} {J. Phys. A}\ }\textbf {\bibinfo {volume} {49}},\ \bibinfo {pages}
		{135001} (\bibinfo {year} {2016})}\BibitemShut {NoStop}%
	\bibitem [{\citenamefont {Borrmann}\ \emph {et~al.}(2000)\citenamefont
		{Borrmann}, \citenamefont {Mulken},\ and\ \citenamefont
		{Harting}}]{Borrmann2000}%
	\BibitemOpen
	\bibfield  {author} {\bibinfo {author} {\bibfnamefont {P.}~\bibnamefont
			{Borrmann}}, \bibinfo {author} {\bibfnamefont {O.}~\bibnamefont {Mulken}}, \
		and\ \bibinfo {author} {\bibfnamefont {J.}~\bibnamefont {Harting}},\
	}\bibfield  {title} {\enquote {\bibinfo {title} {{Classification of Phase
					Transitions in Small Systems}},}\ }\href {\doibase
		10.1103/PhysRevLett.84.3511} {\bibfield  {journal} {\bibinfo  {journal}
			{Phys. Rev. Lett.}\ }\textbf {\bibinfo {volume} {84}},\ \bibinfo {pages}
		{3511} (\bibinfo {year} {2000})}\BibitemShut {NoStop}%
	\bibitem [{\citenamefont {M\"ulken}\ \emph {et~al.}(2001)\citenamefont
		{M\"ulken}, \citenamefont {Borrmann}, \citenamefont {Harting},\ and\
		\citenamefont {Stamerjohanns}}]{Mulken2001}%
	\BibitemOpen
	\bibfield  {author} {\bibinfo {author} {\bibfnamefont {O.}~\bibnamefont
			{M\"ulken}}, \bibinfo {author} {\bibfnamefont {P.}~\bibnamefont {Borrmann}},
		\bibinfo {author} {\bibfnamefont {J.}~\bibnamefont {Harting}}, \ and\
		\bibinfo {author} {\bibfnamefont {H.}~\bibnamefont {Stamerjohanns}},\
	}\bibfield  {title} {\enquote {\bibinfo {title} {Classification of phase
				transitions of finite {B}ose-{E}instein condensates in power-law traps by
				{F}isher zeros},}\ }\href {\doibase 10.1103/PhysRevA.64.013611} {\bibfield
		{journal} {\bibinfo  {journal} {Phys. Rev. A}\ }\textbf {\bibinfo {volume}
			{64}},\ \bibinfo {pages} {013611} (\bibinfo {year} {2001})}\BibitemShut
	{NoStop}%
	\bibitem [{\citenamefont {van Dijk}\ \emph {et~al.}(2015)\citenamefont {van
			Dijk}, \citenamefont {Lobo}, \citenamefont {MacDonald},\ and\ \citenamefont
		{Bhaduri}}]{Dijk2015}%
	\BibitemOpen
	\bibfield  {author} {\bibinfo {author} {\bibfnamefont {W.}~\bibnamefont {van
				Dijk}}, \bibinfo {author} {\bibfnamefont {C.}~\bibnamefont {Lobo}}, \bibinfo
		{author} {\bibfnamefont {A.}~\bibnamefont {MacDonald}}, \ and\ \bibinfo
		{author} {\bibfnamefont {R.~K.}\ \bibnamefont {Bhaduri}},\ }\bibfield
	{title} {\enquote {\bibinfo {title} {Fisher zeros of a unitary {B}ose gas},}\
	}\href {\doibase 10.1139/cjp-2014-0585} {\bibfield  {journal} {\bibinfo
			{journal} {Can. J. Phys.}\ }\textbf {\bibinfo {volume} {93}},\ \bibinfo
		{pages} {830} (\bibinfo {year} {2015})}\BibitemShut {NoStop}%
	\bibitem [{\citenamefont {Gnatenko}\ \emph
		{et~al.}(2017{\natexlab{a}})\citenamefont {Gnatenko}, \citenamefont
		{Kargol},\ and\ \citenamefont {Tkachuk}}]{Gnatenko_2017}%
	\BibitemOpen
	\bibfield  {author} {\bibinfo {author} {\bibfnamefont {K.~P.}\ \bibnamefont
			{Gnatenko}}, \bibinfo {author} {\bibfnamefont {A.}~\bibnamefont {Kargol}}, \
		and\ \bibinfo {author} {\bibfnamefont {V.~M.}\ \bibnamefont {Tkachuk}},\
	}\bibfield  {title} {\enquote {\bibinfo {title} {Time correlation functions
				and {F}isher zeros for q-deformed bose gas},}\ }\href {\doibase
		10.1209/0295-5075/120/30004} {\bibfield  {journal} {\bibinfo  {journal}
			{EPL}\ }\textbf {\bibinfo {volume} {120}},\ \bibinfo {pages} {30004}
		(\bibinfo {year} {2017}{\natexlab{a}})}\BibitemShut {NoStop}%
	\bibitem [{\citenamefont {Gnatenko}\ \emph
		{et~al.}(2017{\natexlab{b}})\citenamefont {Gnatenko}, \citenamefont
		{Kargol},\ and\ \citenamefont {Tkachuk}}]{Gnatenko2017}%
	\BibitemOpen
	\bibfield  {author} {\bibinfo {author} {\bibfnamefont {K.~P.}\ \bibnamefont
			{Gnatenko}}, \bibinfo {author} {\bibfnamefont {A.}~\bibnamefont {Kargol}}, \
		and\ \bibinfo {author} {\bibfnamefont {V.~M.}\ \bibnamefont {Tkachuk}},\
	}\bibfield  {title} {\enquote {\bibinfo {title} {{Two-time correlation
					functions and the Lee-Yang zeros for an interacting Bose gas}},}\ }\href
	{\doibase 10.1103/PhysRevE.96.032116} {\bibfield  {journal} {\bibinfo
			{journal} {Phys. Rev. E}\ }\textbf {\bibinfo {volume} {96}},\ \bibinfo
		{pages} {032116} (\bibinfo {year} {2017}{\natexlab{b}})}\BibitemShut
	{NoStop}%
	\bibitem [{\citenamefont {Biroli}\ and\ \citenamefont
		{Garrahan}(2013)}]{Biroli2013}%
	\BibitemOpen
	\bibfield  {author} {\bibinfo {author} {\bibfnamefont {G.}~\bibnamefont
			{Biroli}}\ and\ \bibinfo {author} {\bibfnamefont {J.~P.}\ \bibnamefont
			{Garrahan}},\ }\bibfield  {title} {\enquote {\bibinfo {title} {Perspective:
				The glass transition},}\ }\href {\doibase 10.1063/1.4795539} {\bibfield
		{journal} {\bibinfo  {journal} {J. Chem. Phys.}\ }\textbf {\bibinfo {volume}
			{138}},\ \bibinfo {pages} {12A301} (\bibinfo {year} {2013})}\BibitemShut
	{NoStop}%
	\bibitem [{\citenamefont {Flindt}\ and\ \citenamefont
		{Garrahan}(2013)}]{Flindt2013}%
	\BibitemOpen
	\bibfield  {author} {\bibinfo {author} {\bibfnamefont {C.}~\bibnamefont
			{Flindt}}\ and\ \bibinfo {author} {\bibfnamefont {J.~P.}\ \bibnamefont
			{Garrahan}},\ }\bibfield  {title} {\enquote {\bibinfo {title} {Trajectory
				{P}hase {T}ransitions, {L}ee-{Y}ang {Z}eros, and {H}igh-{O}rder {C}umulants
				in {F}ull {C}ounting {S}tatistics},}\ }\href {\doibase
		10.1103/PhysRevLett.110.050601} {\bibfield  {journal} {\bibinfo  {journal}
			{Phys. Rev. Lett.}\ }\textbf {\bibinfo {volume} {110}},\ \bibinfo {pages}
		{050601} (\bibinfo {year} {2013})}\BibitemShut {NoStop}%
	\bibitem [{\citenamefont {Hickey}\ \emph {et~al.}(2013)\citenamefont {Hickey},
		\citenamefont {Flindt},\ and\ \citenamefont {Garrahan}}]{Hickey2013}%
	\BibitemOpen
	\bibfield  {author} {\bibinfo {author} {\bibfnamefont {J.~M.}\ \bibnamefont
			{Hickey}}, \bibinfo {author} {\bibfnamefont {C.}~\bibnamefont {Flindt}}, \
		and\ \bibinfo {author} {\bibfnamefont {J.~P.}\ \bibnamefont {Garrahan}},\
	}\bibfield  {title} {\enquote {\bibinfo {title} {Trajectory phase transitions
				and dynamical {L}ee-{Y}ang zeros of the {G}lauber-{I}sing chain},}\ }\href
	{\doibase 10.1103/physreve.88.012119} {\bibfield  {journal} {\bibinfo
			{journal} {Phys. Rev. E}\ }\textbf {\bibinfo {volume} {88}},\ \bibinfo
		{pages} {012119} (\bibinfo {year} {2013})}\BibitemShut {NoStop}%
	\bibitem [{\citenamefont {Hickey}\ \emph {et~al.}(2014)\citenamefont {Hickey},
		\citenamefont {Flindt},\ and\ \citenamefont {Garrahan}}]{Hickey2014}%
	\BibitemOpen
	\bibfield  {author} {\bibinfo {author} {\bibfnamefont {J.~M.}\ \bibnamefont
			{Hickey}}, \bibinfo {author} {\bibfnamefont {C.}~\bibnamefont {Flindt}}, \
		and\ \bibinfo {author} {\bibfnamefont {J.~P.}\ \bibnamefont {Garrahan}},\
	}\bibfield  {title} {\enquote {\bibinfo {title} {Intermittency and dynamical
				{L}ee-{Y}ang zeros of open quantum systems},}\ }\href {\doibase
		10.1103/physreve.90.062128} {\bibfield  {journal} {\bibinfo  {journal} {Phys.
				Rev. E}\ }\textbf {\bibinfo {volume} {90}},\ \bibinfo {pages} {062128}
		(\bibinfo {year} {2014})}\BibitemShut {NoStop}%
	\bibitem [{\citenamefont {Heyl}\ \emph {et~al.}(2013)\citenamefont {Heyl},
		\citenamefont {Polkovnikov},\ and\ \citenamefont {Kehrein}}]{Heyl2013}%
	\BibitemOpen
	\bibfield  {author} {\bibinfo {author} {\bibfnamefont {M.}~\bibnamefont
			{Heyl}}, \bibinfo {author} {\bibfnamefont {A.}~\bibnamefont {Polkovnikov}}, \
		and\ \bibinfo {author} {\bibfnamefont {S.}~\bibnamefont {Kehrein}},\
	}\bibfield  {title} {\enquote {\bibinfo {title} {{Dynamical Quantum Phase
					Transitions in the Transverse-Field Ising Model}},}\ }\href {\doibase
		10.1103/PhysRevLett.110.135704} {\bibfield  {journal} {\bibinfo  {journal}
			{Phys. Rev. Lett.}\ }\textbf {\bibinfo {volume} {110}},\ \bibinfo {pages}
		{135704} (\bibinfo {year} {2013})}\BibitemShut {NoStop}%
	\bibitem [{\citenamefont {Zvyagin}(2016)}]{Zvyagin2016}%
	\BibitemOpen
	\bibfield  {author} {\bibinfo {author} {\bibfnamefont {A.~A.}\ \bibnamefont
			{Zvyagin}},\ }\bibfield  {title} {\enquote {\bibinfo {title} {Dynamical
				quantum phase transitions ({R}eview {A}rticle)},}\ }\href {\doibase
		10.1063/1.4969869} {\bibfield  {journal} {\bibinfo  {journal} {Low Temp.
				Phys.}\ }\textbf {\bibinfo {volume} {42}},\ \bibinfo {pages} {971} (\bibinfo
		{year} {2016})}\BibitemShut {NoStop}%
	\bibitem [{\citenamefont {Heyl}(2018)}]{Heyl2018}%
	\BibitemOpen
	\bibfield  {author} {\bibinfo {author} {\bibfnamefont {M.}~\bibnamefont
			{Heyl}},\ }\bibfield  {title} {\enquote {\bibinfo {title} {Dynamical quantum
				phase transitions: a review},}\ }\href {\doibase 10.1088/1361-6633/aaaf9a}
	{\bibfield  {journal} {\bibinfo  {journal} {Rep. Prog. Phys.}\ }\textbf
		{\bibinfo {volume} {81}},\ \bibinfo {pages} {054001} (\bibinfo {year}
		{2018})}\BibitemShut {NoStop}%
	\bibitem [{\citenamefont {Lamacraft}\ and\ \citenamefont
		{Fendley}(2008)}]{Lamacraft2008}%
	\BibitemOpen
	\bibfield  {author} {\bibinfo {author} {\bibfnamefont {A.}~\bibnamefont
			{Lamacraft}}\ and\ \bibinfo {author} {\bibfnamefont {P.}~\bibnamefont
			{Fendley}},\ }\bibfield  {title} {\enquote {\bibinfo {title} {Order
				{P}arameter {S}tatistics in the {C}ritical {Q}uantum {I}sing {C}hain},}\
	}\href {\doibase 10.1103/PhysRevLett.100.165706} {\bibfield  {journal}
		{\bibinfo  {journal} {Phys. Rev. Lett.}\ }\textbf {\bibinfo {volume} {100}},\
		\bibinfo {pages} {165706} (\bibinfo {year} {2008})}\BibitemShut {NoStop}%
	\bibitem [{\citenamefont {Wei}\ and\ \citenamefont {Liu}(2012)}]{Wei2012}%
	\BibitemOpen
	\bibfield  {author} {\bibinfo {author} {\bibfnamefont {B.-B.}\ \bibnamefont
			{Wei}}\ and\ \bibinfo {author} {\bibfnamefont {R.-B.}\ \bibnamefont {Liu}},\
	}\bibfield  {title} {\enquote {\bibinfo {title} {{Lee-Yang Zeros and Critical
					Times in Decoherence of a Probe Spin Coupled to a Bath}},}\ }\href {\doibase
		10.1103/PhysRevLett.109.185701} {\bibfield  {journal} {\bibinfo  {journal}
			{Phys. Rev. Lett.}\ }\textbf {\bibinfo {volume} {109}},\ \bibinfo {pages}
		{185701} (\bibinfo {year} {2012})}\BibitemShut {NoStop}%
	\bibitem [{\citenamefont {Wei}\ \emph {et~al.}(2014)\citenamefont {Wei},
		\citenamefont {Chen}, \citenamefont {Po},\ and\ \citenamefont
		{Liu}}]{Wei2014}%
	\BibitemOpen
	\bibfield  {author} {\bibinfo {author} {\bibfnamefont {B.-B.}\ \bibnamefont
			{Wei}}, \bibinfo {author} {\bibfnamefont {S.-W.}\ \bibnamefont {Chen}},
		\bibinfo {author} {\bibfnamefont {H.-C.}\ \bibnamefont {Po}}, \ and\ \bibinfo
		{author} {\bibfnamefont {R.-B.}\ \bibnamefont {Liu}},\ }\bibfield  {title}
	{\enquote {\bibinfo {title} {Phase transitions in the complex plane of
				physical parameters},}\ }\href {\doibase 10.1038/srep05202} {\bibfield
		{journal} {\bibinfo  {journal} {Sci. Rep.}\ }\textbf {\bibinfo {volume}
			{4}},\ \bibinfo {pages} {5202} (\bibinfo {year} {2014})}\BibitemShut
	{NoStop}%
	\bibitem [{\citenamefont {Kuzmak}\ and\ \citenamefont
		{Tkachuk}(2019{\natexlab{a}})}]{Kuzmak_2019}%
	\BibitemOpen
	\bibfield  {author} {\bibinfo {author} {\bibfnamefont {A.~R.}\ \bibnamefont
			{Kuzmak}}\ and\ \bibinfo {author} {\bibfnamefont {V.~M.}\ \bibnamefont
			{Tkachuk}},\ }\bibfield  {title} {\enquote {\bibinfo {title} {Detecting the
				{Lee-Yang} zeros of a high-spin system by the evolution of probe spin},}\
	}\href {\doibase 10.1209/0295-5075/125/10004} {\bibfield  {journal} {\bibinfo
			{journal} {{EPL}}\ }\textbf {\bibinfo {volume} {125}},\ \bibinfo {pages}
		{10004} (\bibinfo {year} {2019}{\natexlab{a}})}\BibitemShut {NoStop}%
	\bibitem [{\citenamefont {Kuzmak}\ and\ \citenamefont
		{Tkachuk}(2019{\natexlab{b}})}]{Kuzmak_2019b}%
	\BibitemOpen
	\bibfield  {author} {\bibinfo {author} {\bibfnamefont {A.~R.}\ \bibnamefont
			{Kuzmak}}\ and\ \bibinfo {author} {\bibfnamefont {V.~M.}\ \bibnamefont
			{Tkachuk}},\ }\bibfield  {title} {\enquote {\bibinfo {title} {Probing the
				{Lee-Yang} zeros of a spin bath by correlation functions and entanglement of
				two spins},}\ }\href {\doibase 10.1088/1361-6455/ab3d6b} {\bibfield
		{journal} {\bibinfo  {journal} {J. Phys. B}\ }\textbf {\bibinfo {volume}
			{52}},\ \bibinfo {pages} {205501} (\bibinfo {year}
		{2019}{\natexlab{b}})}\BibitemShut {NoStop}%
	\bibitem [{\citenamefont {Krishnan}\ \emph {et~al.}(2019)\citenamefont
		{Krishnan}, \citenamefont {Schmitt}, \citenamefont {Moessner},\ and\
		\citenamefont {Heyl}}]{Krishnan2019}%
	\BibitemOpen
	\bibfield  {author} {\bibinfo {author} {\bibfnamefont {A.}~\bibnamefont
			{Krishnan}}, \bibinfo {author} {\bibfnamefont {M.}~\bibnamefont {Schmitt}},
		\bibinfo {author} {\bibfnamefont {R.}~\bibnamefont {Moessner}}, \ and\
		\bibinfo {author} {\bibfnamefont {M.}~\bibnamefont {Heyl}},\ }\bibfield
	{title} {\enquote {\bibinfo {title} {Measuring complex-partition-function
				zeros of {I}sing models in quantum simulators},}\ }\href {\doibase
		10.1103/PhysRevA.100.022125} {\bibfield  {journal} {\bibinfo  {journal}
			{Phys. Rev. A}\ }\textbf {\bibinfo {volume} {100}},\ \bibinfo {pages}
		{022125} (\bibinfo {year} {2019})}\BibitemShut {NoStop}%
	\bibitem [{\citenamefont {Binek}(1998)}]{Binek1998}%
	\BibitemOpen
	\bibfield  {author} {\bibinfo {author} {\bibfnamefont {C.}~\bibnamefont
			{Binek}},\ }\bibfield  {title} {\enquote {\bibinfo {title} {{Density of Zeros
					on the Lee-Yang Circle Obtained from Magnetization Data of a Two-Dimensional
					Ising Ferromagnet}},}\ }\href {\doibase 10.1103/PhysRevLett.81.5644}
	{\bibfield  {journal} {\bibinfo  {journal} {Phys. Rev. Lett.}\ }\textbf
		{\bibinfo {volume} {81}},\ \bibinfo {pages} {5644} (\bibinfo {year}
		{1998})}\BibitemShut {NoStop}%
	\bibitem [{\citenamefont {Peng}\ \emph {et~al.}(2015)\citenamefont {Peng},
		\citenamefont {Zhou}, \citenamefont {Wei}, \citenamefont {Cui}, \citenamefont
		{Du},\ and\ \citenamefont {Liu}}]{Peng2015}%
	\BibitemOpen
	\bibfield  {author} {\bibinfo {author} {\bibfnamefont {X.}~\bibnamefont
			{Peng}}, \bibinfo {author} {\bibfnamefont {H.}~\bibnamefont {Zhou}}, \bibinfo
		{author} {\bibfnamefont {B.-B.}\ \bibnamefont {Wei}}, \bibinfo {author}
		{\bibfnamefont {J.}~\bibnamefont {Cui}}, \bibinfo {author} {\bibfnamefont
			{J.}~\bibnamefont {Du}}, \ and\ \bibinfo {author} {\bibfnamefont {R.-B.}\
			\bibnamefont {Liu}},\ }\bibfield  {title} {\enquote {\bibinfo {title}
			{{Experimental Observation of Lee-Yang Zeros}},}\ }\href {\doibase
		10.1103/PhysRevLett.114.010601} {\bibfield  {journal} {\bibinfo  {journal}
			{Phys. Rev. Lett.}\ }\textbf {\bibinfo {volume} {114}},\ \bibinfo {pages}
		{010601} (\bibinfo {year} {2015})}\BibitemShut {NoStop}%
	\bibitem [{\citenamefont {Brandner}\ \emph {et~al.}(2017)\citenamefont
		{Brandner}, \citenamefont {Maisi}, \citenamefont {Pekola}, \citenamefont
		{Garrahan},\ and\ \citenamefont {Flindt}}]{Brandner2017}%
	\BibitemOpen
	\bibfield  {author} {\bibinfo {author} {\bibfnamefont {K.}~\bibnamefont
			{Brandner}}, \bibinfo {author} {\bibfnamefont {V.~F.}\ \bibnamefont {Maisi}},
		\bibinfo {author} {\bibfnamefont {J.~P.}\ \bibnamefont {Pekola}}, \bibinfo
		{author} {\bibfnamefont {J.~P.}\ \bibnamefont {Garrahan}}, \ and\ \bibinfo
		{author} {\bibfnamefont {C.}~\bibnamefont {Flindt}},\ }\bibfield  {title}
	{\enquote {\bibinfo {title} {{Experimental Determination of Dynamical
					Lee-Yang Zeros}},}\ }\href {\doibase 10.1103/physrevlett.118.180601}
	{\bibfield  {journal} {\bibinfo  {journal} {Phys. Rev. Lett.}\ }\textbf
		{\bibinfo {volume} {118}},\ \bibinfo {pages} {180601} (\bibinfo {year}
		{2017})}\BibitemShut {NoStop}%
	\bibitem [{\citenamefont {Fl{\"a}schner}\ \emph {et~al.}(2018)\citenamefont
		{Fl{\"a}schner}, \citenamefont {Vogel}, \citenamefont {Tarnowski},
		\citenamefont {Rem}, \citenamefont {L{\"u}hmann}, \citenamefont {Heyl},
		\citenamefont {Budich}, \citenamefont {Mathey}, \citenamefont {Sengstock},\
		and\ \citenamefont {Weitenberg}}]{Flaschner2018}%
	\BibitemOpen
	\bibfield  {author} {\bibinfo {author} {\bibfnamefont {N.}~\bibnamefont
			{Fl{\"a}schner}}, \bibinfo {author} {\bibfnamefont {D.}~\bibnamefont
			{Vogel}}, \bibinfo {author} {\bibfnamefont {M.}~\bibnamefont {Tarnowski}},
		\bibinfo {author} {\bibfnamefont {B.~S.}\ \bibnamefont {Rem}}, \bibinfo
		{author} {\bibfnamefont {D.-S.}\ \bibnamefont {L{\"u}hmann}}, \bibinfo
		{author} {\bibfnamefont {M.}~\bibnamefont {Heyl}}, \bibinfo {author}
		{\bibfnamefont {J.~C.}\ \bibnamefont {Budich}}, \bibinfo {author}
		{\bibfnamefont {L.}~\bibnamefont {Mathey}}, \bibinfo {author} {\bibfnamefont
			{K.}~\bibnamefont {Sengstock}}, \ and\ \bibinfo {author} {\bibfnamefont
			{C.}~\bibnamefont {Weitenberg}},\ }\bibfield  {title} {\enquote {\bibinfo
			{title} {Observation of dynamical vortices after quenches in a system with
				topology},}\ }\href {\doibase 10.1038/s41567-017-0013-8} {\bibfield
		{journal} {\bibinfo  {journal} {Nat. Phys.}\ }\textbf {\bibinfo {volume}
			{14}},\ \bibinfo {pages} {265} (\bibinfo {year} {2018})}\BibitemShut
	{NoStop}%
	\bibitem [{\citenamefont {Maisi}\ \emph {et~al.}(2014)\citenamefont {Maisi},
		\citenamefont {Kambly}, \citenamefont {Flindt},\ and\ \citenamefont
		{Pekola}}]{Maisi2014}%
	\BibitemOpen
	\bibfield  {author} {\bibinfo {author} {\bibfnamefont {V.~F.}\ \bibnamefont
			{Maisi}}, \bibinfo {author} {\bibfnamefont {D.}~\bibnamefont {Kambly}},
		\bibinfo {author} {\bibfnamefont {C.}~\bibnamefont {Flindt}}, \ and\ \bibinfo
		{author} {\bibfnamefont {J.~P.}\ \bibnamefont {Pekola}},\ }\bibfield  {title}
	{\enquote {\bibinfo {title} {{Full Counting Statistics of Andreev
					Tunneling}},}\ }\href {\doibase 10.1103/physrevlett.112.036801} {\bibfield
		{journal} {\bibinfo  {journal} {Phys. Rev. Lett.}\ }\textbf {\bibinfo
			{volume} {112}},\ \bibinfo {pages} {036801} (\bibinfo {year}
		{2014})}\BibitemShut {NoStop}%
	\bibitem [{\citenamefont {Deger}\ and\ \citenamefont
		{Flindt}(2019)}]{Deger2019}%
	\BibitemOpen
	\bibfield  {author} {\bibinfo {author} {\bibfnamefont {A.}~\bibnamefont
			{Deger}}\ and\ \bibinfo {author} {\bibfnamefont {C.}~\bibnamefont {Flindt}},\
	}\bibfield  {title} {\enquote {\bibinfo {title} {Determination of universal
				critical exponents using {L}ee-{Y}ang theory},}\ }\href {\doibase
		10.1103/PhysRevResearch.1.023004} {\bibfield  {journal} {\bibinfo  {journal}
			{Phys. Rev. Research}\ }\textbf {\bibinfo {volume} {1}},\ \bibinfo {pages}
		{023004} (\bibinfo {year} {2019})}\BibitemShut {NoStop}%
	\bibitem [{\citenamefont {Gaspard}(2012)}]{Gaspard_2012}%
	\BibitemOpen
	\bibfield  {author} {\bibinfo {author} {\bibfnamefont {P.}~\bibnamefont
			{Gaspard}},\ }\bibfield  {title} {\enquote {\bibinfo {title} {Fluctuation
				relations for equilibrium states with broken discrete symmetries},}\ }\href
	{\doibase 10.1088/1742-5468/2012/08/p08021} {\bibfield  {journal} {\bibinfo
			{journal} {J. Stat. Mech.}\ }\textbf {\bibinfo {volume} {2012}},\ \bibinfo
		{pages} {P08021} (\bibinfo {year} {2012})}\BibitemShut {NoStop}%
	\bibitem [{\citenamefont {Friedli}\ and\ \citenamefont
		{Velenik}(2017)}]{friedli_velenik_2017}%
	\BibitemOpen
	\bibfield  {author} {\bibinfo {author} {\bibfnamefont {S.}~\bibnamefont
			{Friedli}}\ and\ \bibinfo {author} {\bibfnamefont {Y.}~\bibnamefont
			{Velenik}},\ }\href {\doibase 10.1017/9781316882603} {\emph {\bibinfo {title}
			{{Statistical Mechanics of Lattice Systems: A Concrete Mathematical
					Introduction}}}}\ (\bibinfo  {publisher} {Cambridge University Press},\
	\bibinfo {year} {2017})\BibitemShut {NoStop}%
	\bibitem [{\citenamefont {Salinas}(2001)}]{Salinas2001}%
	\BibitemOpen
	\bibfield  {author} {\bibinfo {author} {\bibfnamefont {S.~R.~A.}\
			\bibnamefont {Salinas}},\ }\href {\doibase 10.1007/978-1-4757-3508-6} {\emph
		{\bibinfo {title} {Introduction to Statistical Physics}}}\ (\bibinfo
	{publisher} {Springer, New York},\ \bibinfo {year} {2001})\BibitemShut
	{NoStop}%
	\bibitem [{\citenamefont {Kadanoff}(1966)}]{Kadanoff1966}%
	\BibitemOpen
	\bibfield  {author} {\bibinfo {author} {\bibfnamefont {L.~P.}\ \bibnamefont
			{Kadanoff}},\ }\bibfield  {title} {\enquote {\bibinfo {title} {Scaling laws
				for {Ising} models near ${T}_{c}$},}\ }\href {\doibase
		10.1103/PhysicsPhysiqueFizika.2.263} {\bibfield  {journal} {\bibinfo
			{journal} {Phys. Phys. Fiz.}\ }\textbf {\bibinfo {volume} {2}},\ \bibinfo
		{pages} {263} (\bibinfo {year} {1966})}\BibitemShut {NoStop}%
	\bibitem [{\citenamefont {Domb}\ and\ \citenamefont
		{Lebowitz}(1983)}]{Domb1983}%
	\BibitemOpen
	\bibfield  {author} {\bibinfo {author} {\bibfnamefont {C.}~\bibnamefont
			{Domb}}\ and\ \bibinfo {author} {\bibfnamefont {J.~L.}\ \bibnamefont
			{Lebowitz}},\ }\href {\doibase 10.1016/S1062-7901(01)80003-3} {\emph
		{\bibinfo {title} {Phase Transitions and Critical Phenomena}}}\ (\bibinfo
	{publisher} {Academic Press, New York},\ \bibinfo {year} {1983})\BibitemShut
	{NoStop}%
	\bibitem [{\citenamefont {Privman}\ and\ \citenamefont
		{Fisher}(1984)}]{Privman1984}%
	\BibitemOpen
	\bibfield  {author} {\bibinfo {author} {\bibfnamefont {V.}~\bibnamefont
			{Privman}}\ and\ \bibinfo {author} {\bibfnamefont {M.~E.}\ \bibnamefont
			{Fisher}},\ }\bibfield  {title} {\enquote {\bibinfo {title} {Universal
				critical amplitudes in finite-size scaling},}\ }\href {\doibase
		10.1103/PhysRevB.30.322} {\bibfield  {journal} {\bibinfo  {journal} {Phys.
				Rev. B}\ }\textbf {\bibinfo {volume} {30}},\ \bibinfo {pages} {322} (\bibinfo
		{year} {1984})}\BibitemShut {NoStop}%
	\bibitem [{\citenamefont {Privman}(1990)}]{Privman1990}%
	\BibitemOpen
	\bibfield  {author} {\bibinfo {author} {\bibfnamefont {V.}~\bibnamefont
			{Privman}},\ }\href {\doibase 10.1142/1011} {\emph {\bibinfo {title} {Finite
				Size Scaling and Numerical Simulation of Statistical Systems}}}\ (\bibinfo  {publisher} {World
		Scientific Publishing Company},\ \bibinfo {year} {1990})\ p.~\bibinfo {pages}
	{0}\BibitemShut {NoStop}%
	\bibitem [{\citenamefont {Cardy}(1996)}]{Cardy1996}%
	\BibitemOpen
	\bibfield  {author} {\bibinfo {author} {\bibfnamefont {J.}~\bibnamefont
			{Cardy}},\ }\href {\doibase 10.1017/cbo9781316036440} {\emph {\bibinfo
			{title} {Scaling and Renormalization in Statistical Physics}}}\ (\bibinfo
	{publisher} {Cambridge University Press},\ \bibinfo {address} {Cambridge},\
	\bibinfo {year} {1996})\BibitemShut {NoStop}%
	\bibitem [{\citenamefont {Brankov}\ \emph {et~al.}(2000)\citenamefont
		{Brankov}, \citenamefont {Danchev},\ and\ \citenamefont
		{Tonchev}}]{Brankov2000}%
	\BibitemOpen
	\bibfield  {author} {\bibinfo {author} {\bibfnamefont {J.~G.}\ \bibnamefont
			{Brankov}}, \bibinfo {author} {\bibfnamefont {D.~M.}\ \bibnamefont
			{Danchev}}, \ and\ \bibinfo {author} {\bibfnamefont {N.~S.}\ \bibnamefont
			{Tonchev}},\ }\href {\doibase 10.1142/4146} {\emph {\bibinfo {title} {Theory
				of Critical Phenomena in Finite-Size Systems}}}\ (\bibinfo  {publisher}
	{World Scientific},\ \bibinfo {address} {Singapore},\ \bibinfo {year}
	{2000})\BibitemShut {NoStop}%
	\bibitem [{\citenamefont {Kenna}\ and\ \citenamefont
		{Berche}(2014)}]{Kenna2014}%
	\BibitemOpen
	\bibfield  {author} {\bibinfo {author} {\bibfnamefont {R.}~\bibnamefont
			{Kenna}}\ and\ \bibinfo {author} {\bibfnamefont {B.}~\bibnamefont {Berche}},\
	}\bibfield  {title} {\enquote {\bibinfo {title} {{F}isher's scaling relation
				above the upper critical dimension},}\ }\href {\doibase
		10.1209/0295-5075/105/26005} {\bibfield  {journal} {\bibinfo  {journal}
			{{EPL}}\ }\textbf {\bibinfo {volume} {105}},\ \bibinfo {pages} {26005}
		(\bibinfo {year} {2014})}\BibitemShut {NoStop}%
	\bibitem [{\citenamefont {Berche}\ \emph {et~al.}(2012)\citenamefont {Berche},
		\citenamefont {Kenna},\ and\ \citenamefont {Walter}}]{Berche2012}%
	\BibitemOpen
	\bibfield  {author} {\bibinfo {author} {\bibfnamefont {B.}~\bibnamefont
			{Berche}}, \bibinfo {author} {\bibfnamefont {R.}~\bibnamefont {Kenna}}, \
		and\ \bibinfo {author} {\bibfnamefont {J.-C.}\ \bibnamefont {Walter}},\
	}\bibfield  {title} {\enquote {\bibinfo {title} {Hyperscaling above the upper
				critical dimension},}\ }\href {\doibase
		https://doi.org/10.1016/j.nuclphysb.2012.07.021} {\bibfield  {journal}
		{\bibinfo  {journal} {Nucl. Phys. B}\ }\textbf {\bibinfo {volume} {865}},\
		\bibinfo {pages} {115} (\bibinfo {year} {2012})}\BibitemShut {NoStop}%
	\bibitem [{\citenamefont {Binder}(1981)}]{Binder1981}%
	\BibitemOpen
	\bibfield  {author} {\bibinfo {author} {\bibfnamefont {K.}~\bibnamefont
			{Binder}},\ }\bibfield  {title} {\enquote {\bibinfo {title} {{Finite Size
					Scaling Analysis of Ising Model Block Distribution Functions}},}\ }\href
	{\doibase 10.1007/bf01293604} {\bibfield  {journal} {\bibinfo  {journal} {Z.
				Phys. B}\ }\textbf {\bibinfo {volume} {43}},\ \bibinfo {pages} {119}
		(\bibinfo {year} {1981})}\BibitemShut {NoStop}%
	\bibitem [{\citenamefont {Binder}\ and\ \citenamefont
		{Landau}(1984)}]{Binder1984}%
	\BibitemOpen
	\bibfield  {author} {\bibinfo {author} {\bibfnamefont {K.}~\bibnamefont
			{Binder}}\ and\ \bibinfo {author} {\bibfnamefont {D.~P.}\ \bibnamefont
			{Landau}},\ }\bibfield  {title} {\enquote {\bibinfo {title} {Finite-size
				scaling at first-order phase transitions},}\ }\href {\doibase
		10.1103/physrevb.30.1477} {\bibfield  {journal} {\bibinfo  {journal} {Phys.
				Rev. B}\ }\textbf {\bibinfo {volume} {30}},\ \bibinfo {pages} {1477}
		(\bibinfo {year} {1984})}\BibitemShut {NoStop}%
	\bibitem [{\citenamefont {Binder}\ and\ \citenamefont
		{Luijten}(2001)}]{Binder2001}%
	\BibitemOpen
	\bibfield  {author} {\bibinfo {author} {\bibfnamefont {K.}~\bibnamefont
			{Binder}}\ and\ \bibinfo {author} {\bibfnamefont {E.}~\bibnamefont
			{Luijten}},\ }\bibfield  {title} {\enquote {\bibinfo {title} {{Monte Carlo
					tests of renormalization-group predictions for critical phenomena in Ising
					models}},}\ }\href {\doibase 10.1016/s0370-1573(00)00127-7} {\bibfield
		{journal} {\bibinfo  {journal} {Phys. Rep.}\ }\textbf {\bibinfo {volume}
			{344}},\ \bibinfo {pages} {179} (\bibinfo {year} {2001})}\BibitemShut
	{NoStop}%
	\bibitem [{\citenamefont {Huang}(1987)}]{Huang1987}%
	\BibitemOpen
	\bibfield  {author} {\bibinfo {author} {\bibfnamefont {K.}~\bibnamefont
			{Huang}},\ }\href
	{https://www.wiley.com/en-us/Statistical+Mechanics%2C+2nd+Edition-p-9780471815181}
		{\emph {\bibinfo {title} {{Statistical mechanics}}}}\ (\bibinfo  {publisher}
		{John Wiley {\&} Sons, New York},\ \bibinfo {year} {1987})\BibitemShut
		{NoStop}%
		\bibitem [{\citenamefont {Biskup}\ \emph {et~al.}(2000)\citenamefont {Biskup},
			\citenamefont {Borgs}, \citenamefont {Chayes}, \citenamefont {Kleinwaks},\
			and\ \citenamefont {Koteck{\'y}}}]{Biskup2000}%
		\BibitemOpen
		\bibfield  {author} {\bibinfo {author} {\bibfnamefont {M.}~\bibnamefont
				{Biskup}}, \bibinfo {author} {\bibfnamefont {C.}~\bibnamefont {Borgs}},
			\bibinfo {author} {\bibfnamefont {J.~T.}\ \bibnamefont {Chayes}}, \bibinfo
			{author} {\bibfnamefont {L.~J.}\ \bibnamefont {Kleinwaks}}, \ and\ \bibinfo
			{author} {\bibfnamefont {R.}~\bibnamefont {Koteck{\'y}}},\ }\bibfield
		{title} {\enquote {\bibinfo {title} {{General Theory of Lee-Yang Zeros in
						Models with First-Order Phase Transitions}},}\ }\href {\doibase
			10.1103/physrevlett.84.4794} {\bibfield  {journal} {\bibinfo  {journal}
				{Phys. Rev. Lett.}\ }\textbf {\bibinfo {volume} {84}},\ \bibinfo {pages}
			{4794} (\bibinfo {year} {2000})}\BibitemShut {NoStop}%
		\bibitem [{\citenamefont {Janke}\ and\ \citenamefont
			{Kenna}(2001)}]{Janke2001}%
		\BibitemOpen
		\bibfield  {author} {\bibinfo {author} {\bibfnamefont {W.}~\bibnamefont
				{Janke}}\ and\ \bibinfo {author} {\bibfnamefont {R.}~\bibnamefont {Kenna}},\
		}\bibfield  {title} {\enquote {\bibinfo {title} {The {S}trength of {F}irst
					and {S}econd {O}rder {P}hase {T}ransitions from {P}artition {F}unction
					{Z}eroes},}\ }\href {\doibase 10.1023/A:1004836227767} {\bibfield  {journal}
			{\bibinfo  {journal} {J. Stat. Phys.}\ }\textbf {\bibinfo {volume} {102}},\
			\bibinfo {pages} {1211} (\bibinfo {year} {2001})}\BibitemShut {NoStop}%
		\bibitem [{\citenamefont {Touchette}(2009)}]{Touchette2009}%
		\BibitemOpen
		\bibfield  {author} {\bibinfo {author} {\bibfnamefont {H.}~\bibnamefont
				{Touchette}},\ }\bibfield  {title} {\enquote {\bibinfo {title} {The large
					deviation approach to statistical mechanics},}\ }\href {\doibase
			10.1016/j.physrep.2009.05.002} {\bibfield  {journal} {\bibinfo  {journal}
				{Phys. Rep.}\ }\textbf {\bibinfo {volume} {478}},\ \bibinfo {pages} {1}
			(\bibinfo {year} {2009})}\BibitemShut {NoStop}%
		\bibitem [{\citenamefont {Touchette}(2008)}]{Touchette_2008}%
		\BibitemOpen
		\bibfield  {author} {\bibinfo {author} {\bibfnamefont {H.}~\bibnamefont
				{Touchette}},\ }\bibfield  {title} {\enquote {\bibinfo {title} {Simple spin
					models with non-concave entropies},}\ }\href {\doibase 10.1119/1.2794350}
		{\bibfield  {journal} {\bibinfo  {journal} {Am. J. Phys.}\ }\textbf {\bibinfo
				{volume} {76}},\ \bibinfo {pages} {26} (\bibinfo {year} {2008})}\BibitemShut
		{NoStop}%
		\bibitem [{\citenamefont {Touchette}(2010)}]{Touchette_2010}%
		\BibitemOpen
		\bibfield  {author} {\bibinfo {author} {\bibfnamefont {H.}~\bibnamefont
				{Touchette}},\ }\bibfield  {title} {\enquote {\bibinfo {title} {Methods for
					calculating nonconcave entropies},}\ }\href {\doibase
			10.1088/1742-5468/2010/05/p05008} {\bibfield  {journal} {\bibinfo  {journal}
				{J. Stat. Mech.}\ }\textbf {\bibinfo {volume} {2010}},\ \bibinfo {pages}
			{P05008} (\bibinfo {year} {2010})}\BibitemShut {NoStop}%
		\bibitem [{\citenamefont {Alves}\ \emph {et~al.}(2000)\citenamefont {Alves},
			\citenamefont {de~Felicio},\ and\ \citenamefont {Hansmann}}]{Alves2000}%
		\BibitemOpen
		\bibfield  {author} {\bibinfo {author} {\bibfnamefont {N.~A.}\ \bibnamefont
				{Alves}}, \bibinfo {author} {\bibfnamefont {J.~R.~D.}\ \bibnamefont
				{de~Felicio}}, \ and\ \bibinfo {author} {\bibfnamefont {U.~H.~E.}\
				\bibnamefont {Hansmann}},\ }\bibfield  {title} {\enquote {\bibinfo {title}
				{{Partition function zeros and leading-order scaling correction of the 3D
						Ising model from multicanonical simulations}},}\ }\href {\doibase
			10.1088/0305-4470/33/42/302} {\bibfield  {journal} {\bibinfo  {journal} {J.
					Phys. A}\ }\textbf {\bibinfo {volume} {33}},\ \bibinfo {pages} {7489}
			(\bibinfo {year} {2000})}\BibitemShut {NoStop}%
		\bibitem [{\citenamefont {Garc\'{\i}a-Saez}\ and\ \citenamefont
			{Wei}(2015)}]{GarciaSaez2015}%
		\BibitemOpen
		\bibfield  {author} {\bibinfo {author} {\bibfnamefont {A.}~\bibnamefont
				{Garc\'{\i}a-Saez}}\ and\ \bibinfo {author} {\bibfnamefont {T.-C.}\
				\bibnamefont {Wei}},\ }\bibfield  {title} {\enquote {\bibinfo {title}
				{{D}ensity of {Y}ang-{L}ee zeros in the thermodynamic limit from tensor
					network methods},}\ }\href {\doibase 10.1103/PhysRevB.92.125132} {\bibfield
			{journal} {\bibinfo  {journal} {Phys. Rev. B}\ }\textbf {\bibinfo {volume}
				{92}},\ \bibinfo {pages} {125132} (\bibinfo {year} {2015})}\BibitemShut
		{NoStop}%
		\bibitem [{\citenamefont {Xu}\ and\ \citenamefont {del Campo}(2019)}]{Xu2019}%
		\BibitemOpen
		\bibfield  {author} {\bibinfo {author} {\bibfnamefont {Z.}~\bibnamefont
				{Xu}}\ and\ \bibinfo {author} {\bibfnamefont {A.}~\bibnamefont {del Campo}},\
		}\bibfield  {title} {\enquote {\bibinfo {title} {Probing the {F}ull
					{D}istribution of {M}any-{B}ody {O}bservables {B}y {S}ingle-{Q}ubit
					{I}nterferometry},}\ }\href {\doibase 10.1103/PhysRevLett.122.160602}
		{\bibfield  {journal} {\bibinfo  {journal} {Phys. Rev. Lett.}\ }\textbf
			{\bibinfo {volume} {122}},\ \bibinfo {pages} {160602} (\bibinfo {year}
			{2019})}\BibitemShut {NoStop}%
		\bibitem [{\citenamefont {Goldenfeld}(2018)}]{Goldenfeld2018}%
		\BibitemOpen
		\bibfield  {author} {\bibinfo {author} {\bibfnamefont {N.}~\bibnamefont
				{Goldenfeld}},\ }\href {\doibase 10.1201/9780429493492} {\emph {\bibinfo
				{title} {{Lectures on Phase Transitions and the Renormalization Group}}}}\
		(\bibinfo  {publisher} {{CRC} Press, {B}oca {R}aton},\ \bibinfo {year}
		{2018})\BibitemShut {NoStop}%
	\end{thebibliography}
\end{document}